\documentclass[aps,pre,longbibliography,reprint,groupedaddress,floatfix]{revtex4-1}

\usepackage{amsmath}
\usepackage{amssymb}
\usepackage{mathtools}
\usepackage{graphicx,psfrag}
\usepackage{subfig}
\usepackage{verbatim}
\usepackage{import}
\usepackage{grffile}

\begin{document}


\title{Lattice Boltzmann simulation of mixtures with multicomponent van der Waals equation of state}


\author{Kent S. Ridl}
\author{Alexander J. Wagner}
\email[]{alexander.wagner@ndsu.edu}
\homepage[]{www.ndsu.edu/pubweb/$\sim$carswagn}
\affiliation{Department of Physics, North Dakota State University, Fargo, North Dakota 58108, USA}


\date{\today}

\begin{abstract}
We developed a general framework for simulating multicomponent and multiphase systems using the lattice Boltzmann framework.  Despite the fact that there is no restriction on the number of components in principle, in this article we focus an application to two-component mixtures, but we also demonstrate that the algorighm works for larger numbers of components.  To validate our algorithm we separately minimized this underlying free energy to generate theoretical phase diagrams for mixtures of fluids with a van der Waals-like free energy.  All the theoretical phase diagrams are well recovered by our lattice Boltzmann method.
\end{abstract}


\maketitle

\section{Introduction}
In this paper we introduce a lattice Boltzmann (LB) method for multicomponent, multiphase applications.  The development of such methods began shortly after the introduction of lattice Boltzmann methods by McNamara et al. in 1988 \cite{McNamara1988}.  There are three main categories of multiphase and multicomponent models.  The first is based on the lattice gas method by Gunstensen and Rothmann \cite{Gunstensen1991}, and focuses on achieving maximal phase separation of nearly immiscible fluids.  There are some somewhat recent extensions of the LB method based on this approach \cite{ba2016multiple}, but generally it has somewhat fallen out of fashion.

A second approach, developed by Shan and Chen \cite{Shan1993,Shan1995},  is based on mimicking microscopic interaction by introducing a pseudo-potential.  Approaches based on this model continue to be of interest, and developments along these lines are ongoing \cite{PhysRevE.95.043301}.

A third approach, developed by Swift, Orlandini, and Yeomans \cite{Swift1995,Swift1996}, is based on relating the lattice Boltzmann method back to an underlying free energy.  Methods based on this approach continue to be developed, and these approaches are particularly of interest when one can define a free energy functional \cite{LamuraYeomans1999,Semprebon2016}.  

There has been some significant cross-fertilization between the first and second approaches, as equations of state can be selected for pseudo-potential methods.  Also, free energy approaches that originally altered the second moment of the local equilibrium now typically rely on using a mean field forcing approach \citep{LiWagner2007}, although this force is derived from a gradient of a chemical potential rather than an underlying pseudo-potential \cite{Wagner2006,WagnerPooley2007}.

The model presented in this paper relies on deriving chemical potentials from an imposed lattice free energy, and mean field forcing terms are derived from gradients of these chemical potentials.  Here we use the free energy for a mixture of van der Waals fluids as our foundation.  Van der Waals descriptions have received only limited attention recently, since there are free energies with more degrees of freedom that allow for a better fit for specific substances of interest.  However, here we focus on the generic multicomponent, multiphase behavior, and even for a simple mixture of van der Waals fluids with their restricted parameters, a remarkably complex set of phenomena can be recovered.  

Much of this complexity was already understood by van der Waals and co-workers around the turn of the last century.  But the interest of physicists turned to a different direction after that time, and much of this knowledge had been lost in the physics community.  Most modern studies of such fluid mixtures was often restricted to chemistry and chemical process engineering \cite{vanKonynenburg1968,vanKonynenburg1980}, which do not contain descriptions of the phase behavior we observed.  However, we found a recent book by Sengers \cite{Sengers2002, SengersLevelt2002} and several papers by Meijer \cite{Meijer1989, Meijer1990, Meijer1994, Meijer1997, Meijer1999a, Meijer1999b} extremely enlightening.  We then realized that much of our work consisted of re-discovering results that were already known at the beginning of the 20th century.

We demonstrate the ability of our LB approach to recover the phase behavior of a mixture of two van der Waals fluids.  The real interest of using a LB method for describing such a fluid mixture lies in non-equilibrium phenomena.  The development of this method lies in our interest in evaporation phenomena, and the effects that can occur when a change in concentration introduces phase separation fronts \cite{miller2014phase}. One interesting application of a related problem for a mixture of van der Waals fluids by an approach more closely related to the original free energy LB approach looked at the condensation of a gas of two components into a dendritic structure of alternating fluid phases \cite{Pooley2005}.  In this paper, however, we focus first on establishing the appropriateness of our approach to recover the complex phase behavior of these mixtures, since the recovery of equilibrium behavior is a necessary condition of recovering the correct non-equilibrium behavior.

Our paper is structured as follows.  First we introduce a lattice free energy of a mixture of van der Waals fluids in Section \ref{discreteFreesec}, in Section \ref{LBsec} we define our lattice Boltzmann approach, Section \ref{Macroscopidsec} we derive the hydrodynamic limit of our LB approach.  We introduce the simplest possible implementation of this approach in one dimensions in Section \ref{D1Q3sec} and then show some of the most interesting phase diagrams which we recover two ways: the first is by directly minimizing our free energy, and the second is by running a LB simulation.  We show that for a large variety complex phase diagrams the two methods give near identical results, even when we have large density ratios between different phases.  We relate our results back to the common nomenclature of van Konynenburg and Scott \cite{vanKonynenburg1968, vanKonynenburg1980}, which is widely used in chemistry and chemical engineering.  Multiple three-phase regions are recovered, and even metastable regions are recovered.  Alone the elusive four-phase point eluded a recovery by our LB method.  This demonstrates that our lattice Boltzmann method is able to recover complex phase behavior with good accuracy and is a promising candidate to investigate novel non-equilibrium behavior.

\section{\label{discreteFreesec} Discrete Thermodynamics of a Multicomponent System}
The free energy of the familiar one-component van der Waals gas on a lattice can be written ascan be written as
\begin{equation}
  F = \sum_x \left[\rho \theta \ln\left(\frac{\rho}{1-\rho b}\right)-a\rho^2+\frac{\kappa}{2}(\nabla \rho)^2\right]
\end{equation}
where $\rho(x)$ is the local density, $\theta=k_B T$ and $k_B$ is the Boltzmann constant,$\rho b$ is the volumefraction excluded by the repulsive interaction between the particles and $a$ is a parameter representing the attraction between molecules. The $\kappa$ terms is responsible for an interface free energy.

This equation of state predicts a critical point with
\begin{align}
  \theta_{cr} &= \frac{8a}{27b} \label{tetcr}\\
  \rho_{cr} &= \frac{1}{3b} \label{rhocr}\\
  p_c &= \frac{a}{27b^2} \label{pcr}
\end{align}
We later use these relations to express the parameters $a$ and $b$ in terms of the critical temperature and density.

This can be generalized to an multi-component system by using a linear combination of the excluded volumina for each species and quadratic interactions between all components. We introduce a discrete free energy in terms of densities $\rho^c$ of component $c$ for a mixture of van der Waals fluids as:
\begin{align}
F = \sum_x\sum_c &\left[ \rho^c(x)\theta \ln\left( \frac{\rho^c(x)}{1-\sum_{c'}b^{c'} \rho^{c'}(x)} \right) \right. \nonumber \\
& \left. + \sum_{c'}\sum_{\Delta x}\psi^{cc'}(\Delta x) \rho^c(x) \rho^{c'}(x+\Delta x) \right]
\label{discreteFreeEnergy}
\end{align}
Here $b^c$ is a parameter related to the excluded volume of a particle of component $c$, and $\psi^{cc'}$ is an interaction function that includes the strength and the range of the interaction between component $c$ and $c'$.  We assume here that $\psi^{cc'}(x)$ function is symmetric in space $\psi^{cc'}(-x)=\psi^{cc'}(x)$ and in components $\psi^{cc'}(x)=\psi^{c'c}(x)$.

This discrete free energy is equivalent to a standard continuous definition of a van der Waals free energy mixture.  Using a Taylor expansion we can obtain, up to third order derivatives, and interpreting the sum as an integral:
\begin{align}
F = \int dx\sum_c &\Bigg[ \rho^c\theta \ln \left( \frac{\rho^c}{1-\sum_{c'}b^{c'} \rho^{c'}} \right) \nonumber \\
& - \sum_{c'}a^{cc'}\rho^c(x)\rho^{c'} \nonumber \\
& + \left. \frac{1}{2}\sum_{c'}\kappa^{cc'}\nabla\rho^c.\nabla\rho^{c'} \right]
\label{continuousFreeEnergy}
\end{align}
where we have identified the van der Waals attraction and interface parameters in terms of $\psi$ as
\begin{align}
a^{cc'} &= -\sum_{\Delta x}\psi^{cc'}(\Delta x) \label{a-psi} \\
\kappa^{cc'} &= -\sum_{\Delta x}(\Delta x)^2\psi^{cc'}(\Delta x) \label{kappa-psi}
\end{align}
The chemical potential for each component in the mixture is obtained by starting from the discrete free energy (Eq. \ref{discreteFreeEnergy}) where the discrete nature of the free energy replaces the usual functional derivative of the free energy with a simple derivative with respect to $\rho^c(x)$:
\begin{align}
\mu^c(x) =& \frac{\partial F}{\partial \rho^c(x)}\nonumber\\
=& \theta \log\left( \frac{\rho^c(x)}{1-\sum_{c'}b^{c'}\rho^{c'}(x)} \right) + \frac{\theta b^c\rho(x)}{1-\sum_{c'}b^{c'}\rho^{c'}(x)} \nonumber \\
& + 2 \sum_{c'}\sum_{\Delta x} \psi^{cc'}(\Delta x)\rho^{c'}(x+\Delta x)
\label{discreteChemicalPotential}
\end{align}
In general, ensuring bulk equilibrium requires the equality of the chemical potentials and equality of the pressures between the different phases. To ensure thermodynamic consistency, the full chemical potentials are sufficient in the continuous case because of the generalized Gibbs-Duehem relation
\begin{equation}
\nabla_\alpha P_{\alpha\beta} = \sum_c \rho^c \nabla_\beta \mu^c.
\label{GibbsDuehem}
\end{equation}
However, in the discrete case, the validity of a discrete version of the Gibbs-Duehem relation is not guaranteed \cite{Wagner2006}. It is therefore prudent to ensure the consistency by evaluating the equality of the bulk pressure in different phases. We obtain the bulk pressure by assuming constant densities $\rho^c$ in Eq. (\ref{discreteFreeEnergy}). With the assumption that the $\rho^c$ are spatially constant we obtain:
\begin{align}
p =& -\frac{\partial F}{\partial V}=-\sum_c \frac{\partial \mathcal{F}}{\partial \rho^c} \frac{d \rho^c}{dV}=\frac{1}{V}\sum_c \rho^c\frac{\partial \mathcal{F}}{\partial \rho^c}  \nonumber \\
=& \sum_c \left[ \frac{\rho^c\theta}{1-\sum_{c'}b^{c'}\rho^{c'}} - \sum_{c',\Delta x}\psi^{cc'}(\Delta x)\rho^{c}\rho^{c'} \right]  \nonumber \\
=& \sum_c \left[ \frac{\rho^c\theta}{1-\sum_{c'}b^{c'}\rho^{c'}} - \sum_{c'}a^{cc'}\rho^{c}\rho^{c'} \right]
\end{align}
To obtain the gradient terms we have to define a pressure tensor that obeys the Gibbs Duehem relation shown in Eq. (\ref{GibbsDuehem}). This gives
\begin{align}
P_{\alpha\beta} =& \sum_c\left[\frac{\rho^c\theta}{1-\sum_{c'}b^{c'}\rho^{c'}} - \sum_{c'}a^{cc'}\rho^c\rho^{c'} \right. \nonumber \\
&\left. - \sum_{c'}\kappa^{cc'}( \nabla_{\gamma}\rho^c\nabla_{\gamma}\rho^{c'} + \rho^c\nabla^2\rho^{c'} ) \right]\delta_{\alpha\beta}\nonumber\\& + \sum_{cc'}\kappa^{cc'}\nabla_{\alpha}\rho^c\nabla_{\beta}\rho^{c'}
\end{align}
Please note that the bulk values of the pressure for phases that are separated with a flat interface are expected to be identical. For droplets, i.e. phases with a curved interface the pressure inside the drop will be larger, an effect known as Laplace pressure. In this case the divergence of the pressure tensor in Eq. (\ref{GibbsDuehem}) would still be zero. For this equilibrium the chemical potential would also be constant, but the value will be different from the bulk equilibrium value. These interface effects are not studied in the current paper and we restrict our simulations to flat interfaces. Also there will be higher order correction terms to the pressure tensor, as would show up in a fourth order analysis \cite{Wagner2006}.

\section{\label{LBsec} Lattice Boltzmann for a Multicomponent System}
To simulate the dynamics of this multicomponent van der Waals mixture we use a lattice Boltzmann method.  Such a method relies on a discretization of space that we take to conincide with the discrete free energy introduced in the last section.  With each lattice point, we associate a set of lattice velocities $v_i$ that connect it to neighbor lattice points.  

The fundamental variables of the lattice Boltzmann method are densities $f_i(x,t)$ associated with the lattice velocities. The exact interpretation of the $f_i$ in terms of physical quantities remains a little obscure, although there are some recent efforts to shed light on this issue \cite{parsa2017lattice,blommel2018integer}.  At each time step, these densities get moved (streamed) to the lattice point their associated velocity points to.  After this streaming step, the densities at each lattice point get redistributed.  For clarity, the rearranging operation is split into two parts here: a collision operator $\Omega_i^c$ associated with the behavior of ideal gases for one component and a forcing term $F_i^c$ that incorporates the non-ideal interactions as well as momentum exchange among the components.  

For each component we then write a lattice Boltzmann equation:
\begin{equation}
f_i^c(x+v_i, t+1) = f_i^c(x, t) + F_i^c(x, t) + \Omega_i^c(x, t)
\label{LBequation}
\end{equation}
The method conserves the local mass $\rho^c$.  We also define a momentum for each component, $\rho^c u_{\alpha}^c$.  These momenta are not conserved; however, the total local momentum $\rho u_{\alpha}=\sum^c \rho^c u_{\alpha}^c$ is conserved.  Specifically they are defined as
\begin{align}
\rho^c &= \sum_i f_i^c\\
\rho^c u^c_\alpha &= \sum_i f_i^cv_{i\alpha}
\end{align}
where we imply $\rho=\sum_c \rho^c$ and the Greek index $\alpha$ denotes a spatial direction. Here we use the Einstein convention which implies that repeated Greek indices are summed over.

For the collision operator we use the BGK collision operator,
\begin{equation}
\Omega_i^c = \sum_j \Lambda_{ij}(f_j^{c,0} - f_j^c)
\label{BGKcollisionOperator}
\end{equation}
where $f_j^{c,0}$ is the equilibrium distribution associated with velocity $v_j$ for component $c$. The the collision matrix $\Lambda_{ij}$ is diagonal in an appropriate moment basis that contains the hydrodynamic moments and the eigenvalues are inverse relaxation times $\frac{1}{\tau^a}$. This choice implies that the collision operator conserves both the local mass $\rho^c$ and the momentum of each component $\rho^c u^c$. The exchange of momentum between the species in this algorithm is included in the forcing term.

The moments of our equilibrium distribution must be such that the relevant hydrodynamic quantities are recovered:
\begin{align}
\sum_i f_i^{c,0} =& \rho^c \\
\sum_i f_i^{c,0}v_{i\alpha} =& \rho^c u^c_\alpha \\
\sum_i f_i^{c,0}v_{i\alpha}v_{i\beta} =& \rho^c u_{\alpha}^cu_{\beta}^c + \rho^c\theta\delta_{\alpha\beta}
\end{align}
These moments drive the definition of the equilibrium distribution:
\begin{equation}
f_i^{c,0} = \rho^c w_i \left[ 1 + \frac{1}{\theta}v_{i\alpha}u_{\alpha}^c + \frac{1}{2\theta^2}(v_{i\alpha}u_{\alpha}^c)^2 - \frac{1}{2\theta}u_{\alpha}^cu_{\alpha}^c \right]
\label{feq}
\end{equation}
where $w_i$ is a weight associated with a specific lattice velocity $i$.  We note at this point that although $u^c_\alpha$ is not itself a proper hydrodynamic variable, the hydrodynamic mean fluid velocity $u_\alpha$ is composed by weighting the ``velocities" of each component $u^c_\alpha$ by its respective composition $\rho^c/\rho$.

In general forces on component $c$ do not change $\rho^c$, so the zero-order moment of the lattice Boltzmann forcing term $F_i^c$ is
\begin{equation}
\sum_i F_i^c = 0.
\end{equation}
The first velocity moment of $F_i^c$ gives the momentum change of component $c$:
\begin{equation}
\sum_i F_i^cv_{i\alpha} = F_{\alpha}^c.
\end{equation}
The force $F_{\alpha}^c$ has two contributions that combine such that $F_{\alpha}^c = F_{\alpha}^{\mu,c} + F_{\alpha}^{f,c}$:  thermodynamic forcing from chemical potential gradients ($F_{\alpha}^{\mu,c}$) and momentum exchanges from friction between mixture components ($F_{\alpha}^{f,c}$).

As shown in \cite{Wagner2006} the lattice Boltzmann model to this point will contain thermodynamic inconsistencies.  We use the second velocity moment of the forcing term to incorporate corrections ($\Psi_{\alpha\beta}^c$) to the equilibrium behavior:
\begin{equation}
\sum_i F_i^c v_{i\alpha}v_{i\beta} = F_{\alpha}^c u_{\beta} + F_{\beta}^c u_{\alpha} + \Psi_{\alpha\beta}^c.
\end{equation}
Reference \cite{Wagner2006} demonstrates that a fourth-order analysis of forcing methods leads to the following choice for $\Psi_{\alpha\beta}^c$ to ensure consistent thermodynamic equilibrium:
\begin{equation}
\Psi_{\alpha\beta}^c = -\frac{1}{\tau} \left[ (\tau-\frac{1}{4})\frac{F_{\alpha}^cF_{\beta}^c}{\rho^c} + \frac{1}{12}\nabla^2\rho^c \right].
\label{forceCorrectionsPsi}
\end{equation}
The frictional contribution has to be proportional to the velocity difference between the species:
\begin{equation}
F_\alpha^{f,c} = -\sum_{c'} \lambda^{cc'} \frac{\rho^c\rho^{c'}}{\rho^c+\rho^{c'}} (\hat{u}_\alpha^{c'}-\hat{u}_\alpha^c),
\label{forceFriction}
\end{equation}
where the $\hat{u}_\alpha^c$ are true fluid velocities, defined through
\begin{equation}
\hat{u}_\alpha^c = u_\alpha^c +\frac{1}{2\rho^c} F_\alpha^c
\end{equation}
and the minus sign sets the convention that friction acts opposite to any driving forces defined in a positive direction.

A standard lattice Boltzmann method, without extra forcing terms, simulates the evolution of an ideal gas, with an equation of state $p=\rho \theta$. This is consistent with an ideal gas free energy $F^{id}=\sum_x \theta \rho \log(\rho)$, and an ideal chemical potential of $\mu^{id}=\theta \log(\rho)+\theta$.
The conservative force results from the gradient of the non-ideal part of the chemical potential.
\begin{equation}
F_{\alpha}^{\mu,c} = \rho^c\nabla_{\alpha} (\gamma_\mu \mu^{c}-\mu^{c,id}),
\label{forceMuNid}
\end{equation}
where we introduced the factor $\gamma_\mu$ (which can be interpreted as an arbitrary prefactor for the free energy which does not affect the equilibrium behavior) for numerical convenience.
This concludes the brief description of the lattice Boltzmann approach.

\section{\label{Macroscopidsec} Macroscopic equations}
We note that by keeping the form of the lattice Boltzmann equation for a specific component $c$ identical to that of a regular single-component lattice Boltzmann equation, we can sum up our lattice Boltzmann equations over all components ($\sum_c Eq.\ref{LBequation}$) to recover a lattice Boltzmann equation for the entire mixture that also has the single-component form
\begin{equation}
f_i(x+v_i, t+1) - f_i(x, t) + F_i(x, t) = \Omega_i(x, t)
\end{equation}
Given our chosen equilibrium distribution and its moments, we automatically know that the full mixture equations of motion are the standard continuity equation
\begin{equation}
\partial_{t}\rho + \nabla_{\alpha}(\rho \hat{u}_{\alpha}) = 0
\label{discreteContinuity}
\end{equation}
where $\hat{u}_\alpha = u_\alpha +\frac{1}{2\rho} F_\alpha$ and the Navier-Stokes equations
\begin{align}
\partial_{t}(\rho \hat{u}_{\alpha}) +& \nabla_{\beta} (\rho \hat{u}_{\alpha}\hat{u}_{\beta}) = -\nabla_{\alpha}(\rho\theta) + \rho F_{\alpha} + \nonumber \\
& \nabla_{\beta} \left[\tau\rho\theta \left( \nabla_{\beta}\hat{u}_{\alpha} + \nabla_{\alpha}\hat{u}_{\beta} - \frac{2}{3}\nabla_{\gamma}\hat{u}_{\gamma} \delta_{\alpha\beta} \right) \right]
\label{discreteNavierStokes}
\end{align}

However, we must still derive the equations of motion with respect to a single component.  We perform a Taylor expansion of the first term in Eq. (\ref{LBequation}) and make use of Eq. (\ref{BGKcollisionOperator}) to iteratively substitute Eq. (\ref{LBequation}) into itself to obtain an expression for $f_i^c$ in terms of the equilibrium distribution $f_i^{c,0}$
\begin{align}
&\partial_t f_i^{c,0} + v_{i\alpha}\partial_{\alpha}f_i^{c,0} - \tau\partial_t F_i - v_{i\alpha}\partial_{\alpha} F_i + F_i \nonumber \\
&- (\tau-\frac{1}{2})(\partial_t + v_{i\alpha}\partial_{\alpha})_i^2 f_i^{c,0} + O(\partial^3)= \frac{1}{\tau}(f_i^{c,0} - f_i^c)
\label{discreteLBEquation}
\end{align}
Summing over the indices $i$ and using the previous definition of $\hat{u}_{\alpha}^c$ gives 
\begin{equation}
\partial_t \rho^c + \nabla_{\alpha}(\rho^c \hat{u}_{\alpha}^c) = O(\partial^2)
\end{equation}
Since $\hat{u}_{\alpha}^c$ is a function of the non-hydrodynamic component velocity $u_{\alpha}^c$, we wish to eliminate $u^c_{\alpha}$ in favor of the mixture's mean velocity $u_{\alpha}$, which is a hydrodynamic variable.  Defining the component velocity as a deviation from the mean fluid velocity 
\begin{equation}
u^c_{\alpha} = \hat{u}_{\alpha} + \delta u_{\alpha}^c
\label{componentVelocityHydrodynamicDefinition}
\end{equation}
we obtain the component-specific zeroth moment as
\begin{equation}
\partial_t \rho^c + \nabla_{\alpha} \left( \rho^c u^c_{\alpha} \right) = -\nabla_{\alpha}(\rho^c \delta u_{\alpha}^c) + O(\partial^2)
\label{discreteZeroMomentEquation}
\end{equation}

This leaves us with the task of identifying $\delta u_{\alpha}^c$ in terms of the
hydrodynamic quantities. We begin by determining the first velocity moment of Eq. (\ref{discreteLBEquation}) and keeping only first-order terms
\begin{equation}
\partial_t(\rho^c u_{\alpha}^c) + \nabla_{\alpha}(\rho^c u_{\alpha}^c u_{\beta}^c + \rho^c \theta \delta_{\alpha\beta}) + F_{\alpha}^c = O(\partial^2)
\end{equation}
Substituting in the component velocity defined in Eq. (\ref{componentVelocityHydrodynamicDefinition}) and noting that derivatives of the small perturbation $\delta u_{\alpha}^c$ are negligible, we have
\begin{equation}
\partial_t(\rho^c \hat{u}_{\alpha}) + \nabla_{\alpha}(\rho^c \hat{u}_{\alpha} \hat{u}_{\beta}) + \nabla_{\beta}(\rho^c \theta) + F_{\alpha}^c = O(\partial^2)
\end{equation}
Multiplying Eq. (\ref{discreteNavierStokes}) by $\frac{\rho^c}{\rho}$, recognizing that $-\nabla_{\alpha}P_{\alpha\beta} = -\nabla_{\alpha}(\rho\theta)+F_{\alpha}$, and absorbing second-order terms into $O(\partial^2)$ allows a substitution for the first 2 terms
\begin{equation}
-\frac{\rho^c}{\rho}\nabla_{\alpha}P_{\alpha\beta} + \nabla_{\beta}(\rho^c \theta) + F_{\alpha}^c = O(\partial^2)
\end{equation}
Finally, we substitute in the Gibbs-Duhem relation (Eq. \ref{GibbsDuehem}) and the force definitions from Eqs. (\ref{forceFriction}) and (\ref{forceMuNid}) to obtain
\begin{align}
& -\frac{\rho^c}{\rho}\sum_{c'} \rho^{c'} \nabla_\beta \gamma_\mu \mu^{c'} + \nabla_{\beta}(\rho^c \theta) + \rho^c\nabla_{\alpha} (\gamma_\mu \mu^{c}-\mu^{c,id}) \nonumber \\
& - \sum_{c'} \lambda^{cc'} \frac{\rho^c\rho^{c'}}{\rho^c+\rho^{c'}} (\hat{u}_\alpha^{c'}-\hat{u}_\alpha^c) = O(\partial^2)
\end{align}
Using $\theta\rho^c\nabla\log(\rho^c)=\theta\nabla\rho^c$, the second term above is recognized as an ideal chemical potential gradient (for an isothermal system), which cancels with part of the non-ideal chemical potential driving force. Given the definition of the full density $\rho$, we can simplify the expression as
\begin{equation}
\sum_{c'} \frac{\rho^c\rho^{c'}}{\rho} \gamma_\mu \nabla_{\alpha}(\mu^c-\mu^{c'}) - \sum_{c'} \lambda^{cc'} \frac{\rho^c\rho^{c'}}{\rho} (\hat{u}_\alpha^{c'}-\hat{u}_\alpha^c) = O(\partial^2)
\end{equation}
Expanding this equation and and substituting in the Gibbs-Duhem relation allows us to rewrite as
\begin{equation}
-\gamma_\mu \nabla_{\alpha} \mu^c = \sum_{c'} \lambda^{cc'} \frac{\rho^{c'}}{\rho} (\hat{u}_\alpha^c-\hat{u}_\alpha^{c'}) - \frac{1}{\rho}\nabla_{\beta}P_{\alpha\beta} + O(\partial^2)
\label{MaxwellStefan}
\end{equation}
This linear system of equations is the definition of Maxwell-Stefan diffusion with the addition of a baro\~diffusion term $\nabla_{\beta}P_{\alpha\beta}$ that captures an average pressure gradient force acting on each component.  Provided the pressure tensor is non-singular, one may use the hydrodynamic substitution for $u_{\alpha}^c$ (Eq. \ref{componentVelocityHydrodynamicDefinition}), and in principle, Eq. (\ref{MaxwellStefan}) will always yield a general solution for $\delta u_{\alpha}^c$. 

In the specific case of symmetric, constant $\lambda^{cc'} = \lambda$ and a divergence-free pressure tensor - such as for a two-component simulation - this reduces to general Fickian diffusion where
\begin{equation}
\delta u^c = -\frac{1}{\lambda} \gamma_\mu\partial_{\alpha} \mu^c + O(\partial^2)
\end{equation}
and the term $1/\lambda$ is identified as the component mobility.

\section{\label{D1Q3sec} D1Q3 Implementation}
The simplest implementation of this lattice Boltzmann method, and one entirely sufficient to recover the phase-behavior, consists of a one dimensional model with only three velcoities $v_i \in \{0, +1, -1\}$.  Given the model definition in the previous section, the lattice Boltzmann equations for each component and velocity at a given lattice site are explicitly:
\begin{align}
f_0^c(t+1) \mathrel{+}=& \frac{1}{\tau} \left( \rho^c-\rho^c\theta-\rho^c u^{c2}-f_0^c(t) \right) \nonumber \\
& - (2F^c u^c - \Psi^c) \\
f_{+1}^c(t+1) \mathrel{+}=& \frac{1}{\tau} \left[ \frac{1}{2}(\rho^c u^{c2}+\rho^c u^c+\rho^c\theta-f_{+1}^c(t)) \right] \nonumber \\
& - \left( -F^c u^c - \frac{1}{2}F^c + \frac{1}{2}\Psi^c \right) \\
f_{-1}^c(t+1) \mathrel{+}=& \frac{1}{\tau} \left[ \frac{1}{2}(\rho^c u^{c2} - \rho^c u^c + \rho^c\theta - f_{-1}^c(t)) \right] \nonumber \\
& - \left( -F^c u^c + \frac{1}{2}F^c + \frac{1}{2}\Psi^c \right)
\end{align}
Eqs. (\ref{a-psi}) and (\ref{kappa-psi}) can also be explicitly expanded in terms of the velocity set for the D1Q3 model, with the velocities implicitly corresponding to $\Delta x$ for a single time step. This allows us to identify the parts of $\psi^{cc'}$, assuming we use a support of only the central lattice point and its neighbors.  Beginning with Eq. (\ref{kappa-psi}), we have:
\begin{align}
\kappa^{cc'} =& -\psi^{cc'}(x+1)-\psi^{cc'}(x-1)
\label{kappa-psi-D1Q3}
\end{align}
Eq. (\ref{a-psi}) expands to
\begin{align}
a^{cc'} =& -\psi^{cc'}(x)-\psi^{cc'}(x+1)-\psi^{cc'}(x-1) \nonumber \\
=& -\psi^{cc'}(x) + \kappa^{cc'}
\label{a-psi-D1Q3}
\end{align}
Eqs. (\ref{kappa-psi-D1Q3}) and (\ref{a-psi-D1Q3}) taken together imply the complete set of $\psi^{cc'}$
\begin{align}
\psi^{cc'}(x) =& \kappa^{cc'} - a^{cc'} \\
\psi^{cc'}(x+1) = \psi^{cc'}(x-1) =& -\frac{1}{2}\kappa^{cc'}
\end{align}

Expansion of the interaction term in Eq. (\ref{discreteFreeEnergy}) and substituting in Eqs. (\ref{kappa-psi-D1Q3}) and (\ref{a-psi-D1Q3}) allows the identification of an appropriate gradient stencil for the model.  In this particular model, we recover the standard second-order finite difference Laplace stencil.

To aid the accuracy and help control the stability of the lattice Boltzmann simulations in the regions where we anticipated phase separation with sizeable density ratios and to accelerate convergence, we implemented two strategies.  First, lattice Boltzmann simulations were initialized with $tanh$ density profiles with bulk values equal to the results of our free energy minimization.  And second, we sought to initialize the density profiles with an interface width, $w$, that was close to the equilibrium interface width.  

The initial interface width we imposed drove the choice of two simulation parameters: $\kappa^{cc'}$ and a chemical potential coefficient $\gamma_{\mu}$ of eqn. (\ref{forceMuNid}), which is a numerical parameter used to control the abruptness with which the chemical potential-based forcing $F^{\mu,c}$ is applied.  The relationship of these parameters to the interface width was determined in \cite{WagnerPooley2007} to be:
\begin{equation}
w(\kappa,\theta / \theta_{cr}) = \sqrt{\frac{2\kappa^{cc'}}{\frac{\theta_{cr}}{\theta}-1}}
\end{equation} 
where $\theta_{cr}$ is the critical temperature for a component and $\theta$ is the isothermal lattice temperature, and setting
\begin{equation}
c_s = \sqrt{1-4\kappa^{cc'}\rho_l\gamma_{\mu}}
\label{kappaGammaMuRelation}
\end{equation} 
where $c_s$ is the lattice speed of sound (equal to $\frac{1}{\sqrt{3}}$) and $\rho_l$ is the expected liquid density determined by minimizing the free energy. Since this was derived for a single-component system, this is only an approximate guide for the multicomponent interface width.

Finally, we tested two implementations of thermodynamic forcing ($F_{\alpha}^{\mu,c}$):
\begin{enumerate}
  \item Non-ideal chemical potential gradient of component $c$ 
  \begin{equation}
  F_{\alpha}^{\mu,c} = \rho^c\nabla^D_{\alpha} (\gamma_\mu \mu^{c}- \theta ln \rho^c)
  \label{forceMuId}
  \end{equation}
  where $\nabla^D$ corresponds to a central-difference discretization of the gradient operator. 
  This is the choice that naively corresponds to the thermodynamic prediction of the force.  This method is termed the ``nid" method.
  \item Gradient of the chemical potential of component $c$ less the ideal pressure from component $c$ 
  \begin{equation}
  F_{\alpha}^{\mu,c} = \rho^c\nabla^D_{\alpha}\gamma_\mu \mu^c - \theta\nabla^D_{\alpha}\rho^c
  \label{forceMuLog}
  \end{equation}
  This method is termed the ``log" method.
\end{enumerate}
Both are expressions for the non-ideal chemical potential driving force in Eq. (\ref{forceMuNid}); however, the numeric representations of each yield results that differ in both accuracy and stability, as will be seen in the following section.

\section{Verification for Mixtures of van der Waals Fluids}
In this section we will examine the ability of our lattice Boltzmann method to recover the complex phase behavior of mixtures of van der Waals fluids. The first, quite extensive, section regards two-component mixtures. We examine the key types of phase diagrams obtainable for such mixtures. A smaller second part shows a single example of a three-component van der Waals fluid, showing four phase coexistence.

\subsection{Two van der Waals Fluids}
To verify the theory in the preceding section, we focused on the simplest case of a mixture of two VDW fluids.  We chose to specify the two components of the mixture via three degrees of freedom: component A and B critical temperatures ($\theta_{cr}^A, \theta_{cr}^B$) and component B critical density ($\rho_{cr}^B$); we fix the component A critical density $\rho_{cr}^A = 1$.  The other properties of the components were determined by the VDW relations using eqns. (\ref{tetcr}) and (\ref{rhocr}):
\begin{align}
  a^{A} &= \frac{9(\theta_{cr}^A)^2}{8 \rho_{cr}^A} & b^A &= \frac{1}{3\rho_{cr}^A}\\
  a^{B} &= \frac{9(\theta_{cr}^B)^2}{8 \rho_{cr}^B} & b^B &= \frac{1}{3\rho_{cr}^B}
\end{align}
 The energetic interaction between components A and B was controlled by the implementation of a geometric mixing rule applied to $a^A$ and $a^B$ \cite{vanLaar1904,Meijer1989,Benmekki1987}:
\begin{equation}
a^{AB} = \nu \sqrt{a^Aa^B}
\label{vdwConstantAB}
\end{equation}
with the parameter $\nu$ allowing the interaction to deviate from the geometric mixing rule.  With this rule in place, a neutral interaction corresponds to $\nu = 1$, a repulsive interaction corresponds to $\nu < 1$, and an attractive interaction corresponds to $\nu > 1$.  Note that we also used the same parameter $\nu$ to control the cross-component interactions in the interface terms of the chemical potential (i.e. $\kappa^{cc'} = \nu \kappa^{cc}$ for $c \neq c'$ and $c \in {A,B}$).

\begin{figure}
  \centering
\resizebox{\columnwidth}{!}{\input{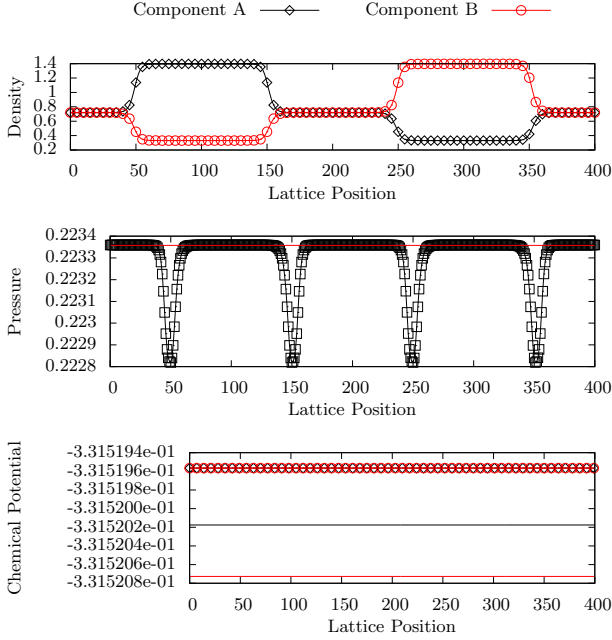}}
\caption{Simulation results from our baseline phase diagram (see Figure \ref{graphBaselineCase}) for the initial $(\rho^A,\rho^B)$ pair (0.8, 0.8), which exhibits thermodynamically consistent 3-phase equilibrium (see text for details).}
\label{graphBaselineNa008Nb008}
\end{figure}

An illustration of our LB simulations is shown in Figure \ref{graphBaselineNa008Nb008}.  This Figure shows the density profile along with the associated pressure and chemical potentials that the LB simulation recovers for the $(\rho^A,\rho^B)$ pair (0.8, 0.8) for $\theta_{cr}^A=\theta_{cr}^B=0.4$ (note that $\theta=1/3$ in all LB simulations), $\rho_{cr}^A=\rho_{cr}^B=1, \nu=0.5$, as well as $\kappa^{c,c'}=0.1$ and $\gamma_\mu$ is given by Eq. (\ref{kappaGammaMuRelation}). This density pair lies in the middle of the three-phase region in Figure \ref{graphBaselineCase}.  The simulation is run for 50,000 iterations. The density profile shows an $A$-rich and a $B$-rich liquid domain both of which  are separated by two gas-domains. From these domains we obtain the compositions of the three phases. The chemical potentials are constant across the lattice and the pressure in the bulk phases is also constant. This confirms that the simulation has recovered the thermodynamic equilibrium.  The simulated bulk pressure converges to a constant value across the lattice to better than $10^{-11}$, and it matches theoretical expectations to approximately $10^{-6}$.  There are much larger discrepancies at the interfaces which we attribute to higher order gradient terms in the pressure, that were not investigated for the current paper. Note that the equality of the bulk pressure and chemical potential are sufficient to ensure the correctness of the phase diagrams.

The simulated chemical potentials for both components are constant across the entire lattice - interfaces included - to better than $10^{-12}$, and the values also both match theoretical expectations to approximately $10^{-6}$ (we note here that our free energy minimization routine converges with an uncertainty of order $10^{-6}$).  Finally, we note that both the pressure and chemical potentials converge to near-machine accuracy constant values if the simulations are allowed sufficient time to run.

To extensively test the equilibrium behavior of our method, we created phase diagrams for several two-component mixtures spanning a range of component properties to which we could compare a series of LB simulations.  All of the LB simulations are isothermal with conserved mean densities, so a natural way to present our data is to plot coexistence curves on a density-density plane.  Note that every point on this plane represents a mixture with its own equilibrium pressure. 

\begin{figure}
\centering
\includegraphics[width=\columnwidth]{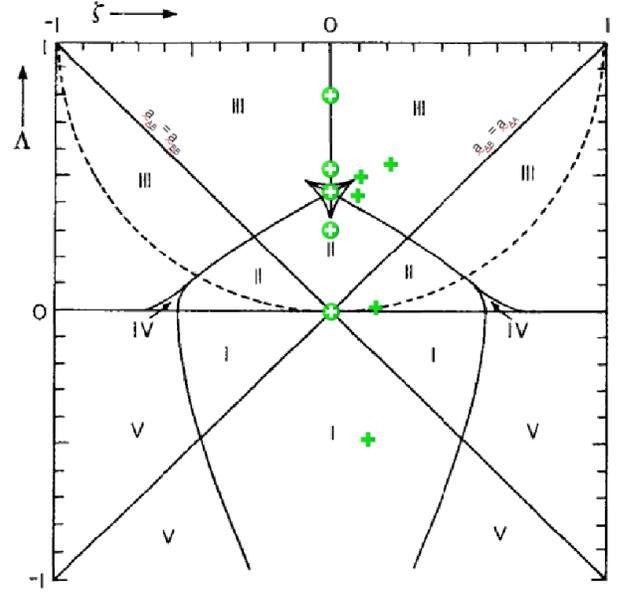}
\caption{The global phase diagram for a binary van der Waals fluid mixture shows five regions (I-V) reproducible by the VDW EOS with cross symbols approximating the state of phase diagrams depicted in this paper.  The 5 open symbols with a circular background indicate symmetric mixtures with equal molecular sizes for each component ($\xi = 0$), and the 5 solid symbols indicate asymmetric mixtures for unequal molecular sizes.  The dashed curve depicts neutral cross-component interactions ($\nu=1$) for the geometric mixing rule in Eq. (\ref{vdwConstantAB}).  Azeotropy is not relevant to the current study and is not depicted here.  Adapted from \cite{Scott1987}.}
\label{globalPhaseDiagram}
\end{figure}

Scott and van Konynenburg \cite{vanKonynenburg1968,vanKonynenburg1980} developed a taxonomy of binary van der Waals mixtures. It classifies phase behavior based on characteristics of pressure-temperature phase diagrams. The basic idea is that the van der Waals mixtures are characterized by three dimensionless parameters
\begin{align}
  \xi &= \frac{b^B - b^A}{b^A + b^B} =\frac{\rho_{cr}^A-\rho_{cr}^B}{\rho_{cr}^A+\rho_{cr}^B}\\
  \zeta &= \left(\frac{a^B}{(b^B)^2}-\frac{a^A}{(b^A)^2}\right)\bigg/\left(\frac{a^A}{(b^A)^2}+\frac{a^B}{(b^B)^2}\right) \nonumber\\
  &=\frac{p_{cr}^B-p_{cr}^A}{p_{cr}^A+p_{cr}^B}\\
\Lambda &= \left(\frac{a^A}{(b^A)^2}-\frac{2a^{AB}}{b^Ab^B}+\frac{a^B}{(b^B)^2}\right)\bigg/\left(\frac{a^A}{(b^A)^2}+\frac{a^B}{(b^B)^2}\right)
\end{align}
which are obtained from the 5 parameters in the free energy (\ref{continuousFreeEnergy}) by using the freedom to choose a time and length scale.
 $\xi$ characterizes the relative size of the constituent components, with equal sizes corresponding to $\xi = 0$. $\zeta$ is a measure of the asymmetry of the critical pressures of the pure components (\ref{pcr}), and $\Lambda$ indicates whether the $A-B$ interactions are more attractive ($\Lambda<0$) or less attractive ($\Lambda>0$) than that of a neutral mixture. In particular binary fluid phase separation is only possible for $\Lambda>0$. 

Figure \ref{globalPhaseDiagram} has been adapted from \cite{Scott1987}. This diagram shows different classes of P-T phase diagrams as detailed in their paper. We include it here to outline where our phase diagrams, presented later, fall within the context of their classification.

Figure \ref{globalPhaseDiagram} only shows the classification for $\xi=0$. We see that it is symmetric with respect to the $\zeta = 0$ vertical axis.  For non-zero values of $\xi$, the vertical axis shifts left or right and the regions for each phase diagram type compress or expand accordingly.  However, the relationships among phase diagram types remain the same.  In our case specifically, the values of $\xi$ given by our parameter choices were $-0.1 <= \xi <= 0$, which has a negligible effect on the layout of Figure \ref{globalPhaseDiagram}.

At this point we note that although we were able to sample a variety of $\Lambda$ values, our ability to sample a wide array of $\zeta$ values was limited.  This was due to the dependence of the $a^{cc'}$ and $b^c$ parameters on the critical temperatures of the components.  Moderate to large values of $\zeta$ drove the selection of critical temperatures that very quickly lead to numerical instabilities in the lattice Boltzmann method.  We were able to remedy these instabilities by application of stabilization methods outlined later in this section, but low values of $\zeta$ were the only ones that were able to sample reliably without manual intervention.    

To obtain a theoretical density-density phase diagram from the free energy (\ref{continuousFreeEnergy}) we sampled combinations of $(\rho^A,\rho^B)$ for given parameters. The process is explained in more detail in Appendix \ref{App1}.  It turned out to be necessary to perform stability analysis for each point to ascertain whether a mixture is unstable and what variation of the density will lead to a reduction of the free energy.  We obtain the 2x2 Hessian $\mathbf{H}$ of free energy derivatives
\begin{equation}
\mathbf{H} = 
\begin{bmatrix}
\frac{\partial^2\psi}{\partial(\rho^A)^2} & \frac{\partial^2\psi}{\partial\rho^B \partial\rho^A} \\
\frac{\partial^2\psi}{\partial\rho^A\partial\rho^B} & \frac{\partial^2\psi}{\partial(\rho^B)^2}
\end{bmatrix}
\end{equation}
If the determinant of $\mathbf{H}$ at point (A,B) was negative, a single negative eigenvalue of the Hessian exists and phase separation was to be expected at that point.  We then numerically minimized the free energy given by Eq. (\ref{discreteFreeEnergy}) at that point while allowing for three co-existing phases.  The eigenvector of the negative eigenvalue from the stability analysis was used to set the direction of the first step for the minimization.  The results of the free energy minimization were logged, allowing us to define both binodal lines and a spinodal region for the phase diagram.  In all graphs that follow, binodal lines are all depicted by solid black lines, and the edges of the light gray regions approximate the spinodal regions (referred to as quasi-spinodals) where the fluid will be unconditionally unstable towards a phase separation.

\begin{figure}
\begin{center}
\resizebox{\columnwidth}{!}{\input{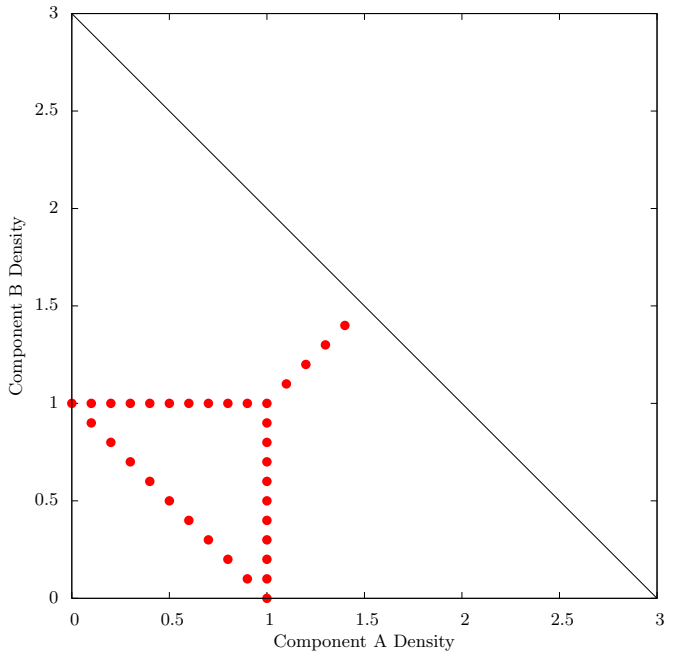}}
\end{center}
\caption{This figure shows the set of points used to initialize the LB simulations to test the two-phase regions of our phase diagrams.  The algorithm moves vertically from the A-component axis, horizontally from the B-component axis, diagonally from axis to axis, and diagonally from (1.0,1.0) to (1.4,1.4).  The point (1.5,1.5) is also tested if a mixture components are asymmetric enough to admit it.}
\label{twoPhaseTestPoints}
\end{figure}

The LB simulations were initialized with a range of $(\rho^A,\rho^B)$ density pairs in a near-equilibrium profile and allowed to iterate for 50,000 time steps.  We selected up to 35 density pairs shown in Figure \ref{twoPhaseTestPoints} to test in regions of the theoretical phase diagram that were anticipated to exhibit 2-phase behavior; anticipated 3-phase regions were exhaustively tested.  The densities associated with the resulting phases were logged when each simulation concluded.  Simulations covered a range of critical temperature, critical density, and interaction parameters to test a variety of phase diagram structures.

The lattice Boltzmann simulations that tested each phase diagram were automated to consistently and comprehensively test all two- and three-phase regions.   To cover the two-phase regions, we selected ``paths" through the phase diagrams that were general enough to ensure at least one lattice Boltzmann simulation would occur in all of the anticipated regions: vertical from the A-component axis (1,0) to (1,1), horizontal from the B-component axis (0,1) to (1,1), diagonal from axis-to-axis (1,0) to (0,1), and diagonal from (1,1) to near the van der Waals singularity line at (1.4, 1.4). These path are shown in Figure \ref{twoPhaseTestPoints}.  Note that occasionally the asymmetry of the components in a mixture will also include the point (1.5, 1.5).

In general, we chose $\kappa^{cc'} = 0.1$ for our LB simulations.  The major exception to this is in simulating the 2-phase behavior in the binary liquid regions of a phase diagram, where we allowed $\kappa^{cc'}$ to linearly increase from 0.15 at (1.0,1.0) to 0.5 at (1.4,1.4).  Deviations from these values are noted in the captions of the associated phase diagrams.

Starting with a specified value for $\kappa^{cc'}$, all simulations began by estimating the width of the equilibrium interface given in \cite{WagnerPooley2007} (for a single component) as the minimum interface width
\begin{equation}
w_{min} = \frac{1}{\sqrt{4\rho_v|\theta_{cr}-\theta|}}
\end{equation}
which we modified to allow phase diagrams with components to be warmer than their respective critical temperatures.  This initial width from the single-component theory in \cite{WagnerPooley2007} proved to overestimate the equilibrium interface width in most cases.  This had the effect of shifting the resulting bulk density values, affecting the accuracy of the simulations.  To improve the accuracy, this initial estimate was then iterated to an equilibrium state (usually 50,000 time steps) where the equilibrium interface width was numerically measured.  The measured interface width was used to re-initialize the simulation, and a coefficient was calculated to preserve the relationship $w \propto \sqrt{\kappa^{cc'}}$.  The value of $\gamma_{\mu}$ was initialized according to the relation in Eq. (\ref{kappaGammaMuRelation}).  When a simulation was unstable ($\sim5\%$ of the total simulations), $\gamma_{\mu}$ was numerically optimized to find the maximum value that would provide a stable simulation.  If this automation failed to find a stable simulation, parameters were manually tuned by either decreasing $\gamma_{\mu}$ by a factor of anywhere from 2-10 or setting the initial interface width to 2 lattice spaces.  This occurred 13 times in $\sim$160 simulations in the two-phase regions, and not at all in $\sim$5600 simulations in the three-phase regions.

\begin{figure}
\resizebox{\columnwidth}{!}{\input{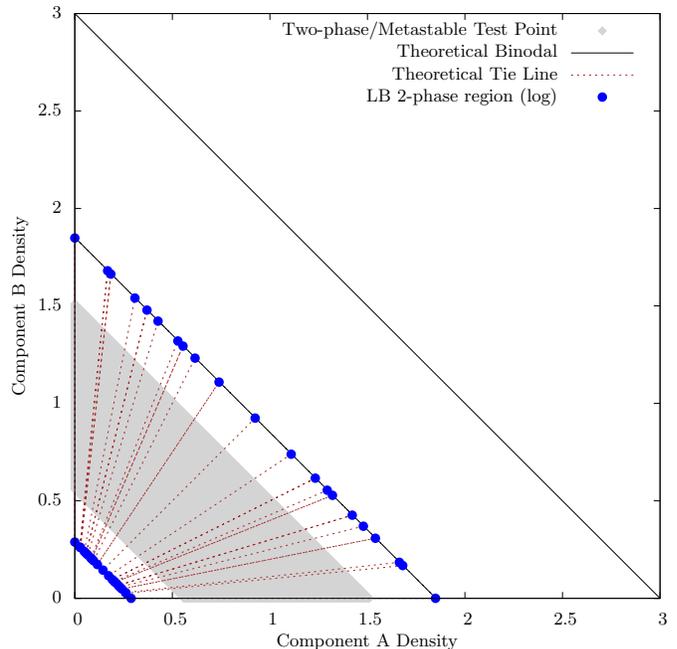}}
\caption{This figure shows the phase diagram for a van der Waals mixture of two identical components.  It was generated using parameters of $\theta_{cr}^A = \theta_{cr}^B = 0.4$, $\rho_{cr}^B = 1.0$, and a neutral interaction parameter $\nu = 1.0$ $(\xi=0.0, \zeta=0.0, \Lambda=0.0)$.  Overlaid on top of the theoretical diagram generated by free energy minimization are the results of the LB simulations.  Also depicted is a diagonal connecting the VDW equation discontinuities for both components.}
\label{graphTwoIdenticalComponents}
\end{figure}

The subsections that follow outline our LB simulation results for the phase diagrams indicated in Figure \ref{globalPhaseDiagram}.  Results are grouped in three ways: symmetric components (solid symbols in Figure \ref{globalPhaseDiagram}), asymmetric components (open symbols in Figure \ref{globalPhaseDiagram}), and the so-called ``shield" region (enclosed region around $\zeta = 0,\Lambda = 0.4364$ in Figure \ref{globalPhaseDiagram}).  Several LB simulations touched areas of numerical instability, but the automated parameter tuning performed well and lattice Boltzmann simulations were still able to reproduce all binodals.

\begin{figure}
\centering
\resizebox{\columnwidth}{!}{\input{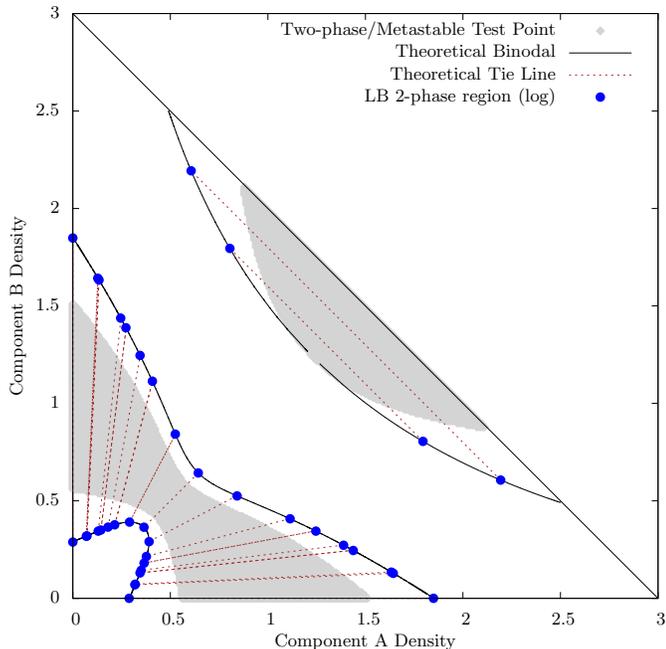}}
\caption{This Type II phase diagram is identical to the de-facto single component mixture except phase separation was induced by setting $\nu = 0.7$ $(\xi=0.0, \zeta=0.0, \Lambda=0.3)$.  The interaction parameter is not quite repulsive enough to connect all regions of phase separation or provoke 3-phase behavior.}
\label{graphModerateRepulsive}
\end{figure}

\subsubsection{Symmetric Components}
The first test case was for that of a mixture of two identical components with neutral interactions. Figure \ref{graphTwoIdenticalComponents} shows the phase diagram recovered by minimizing the free energy of such a system along with the associated lattice Boltzmann simulations.  As we expected for a de-facto single-component simulation, we obtained a perfectly symmetrical diagram with straight binodal lines connecting equal densities on the A- and B-component axes.  The values of the phase-separated densities on each axis corresponded to the results of a single-component simulation given the same initial conditions. The only addition to a single-component system is the entropy of mixing which ensures that the liquid and the gas have identical compositions, which is indicated here by the fact that the tie-lines for liquid-gas coexistence would all meet in the origin.

\begin{figure*}
\captionsetup[subfloat]{captionskip=10pt}
\centering
\subfloat[][]{\resizebox{\columnwidth}{!}{\input{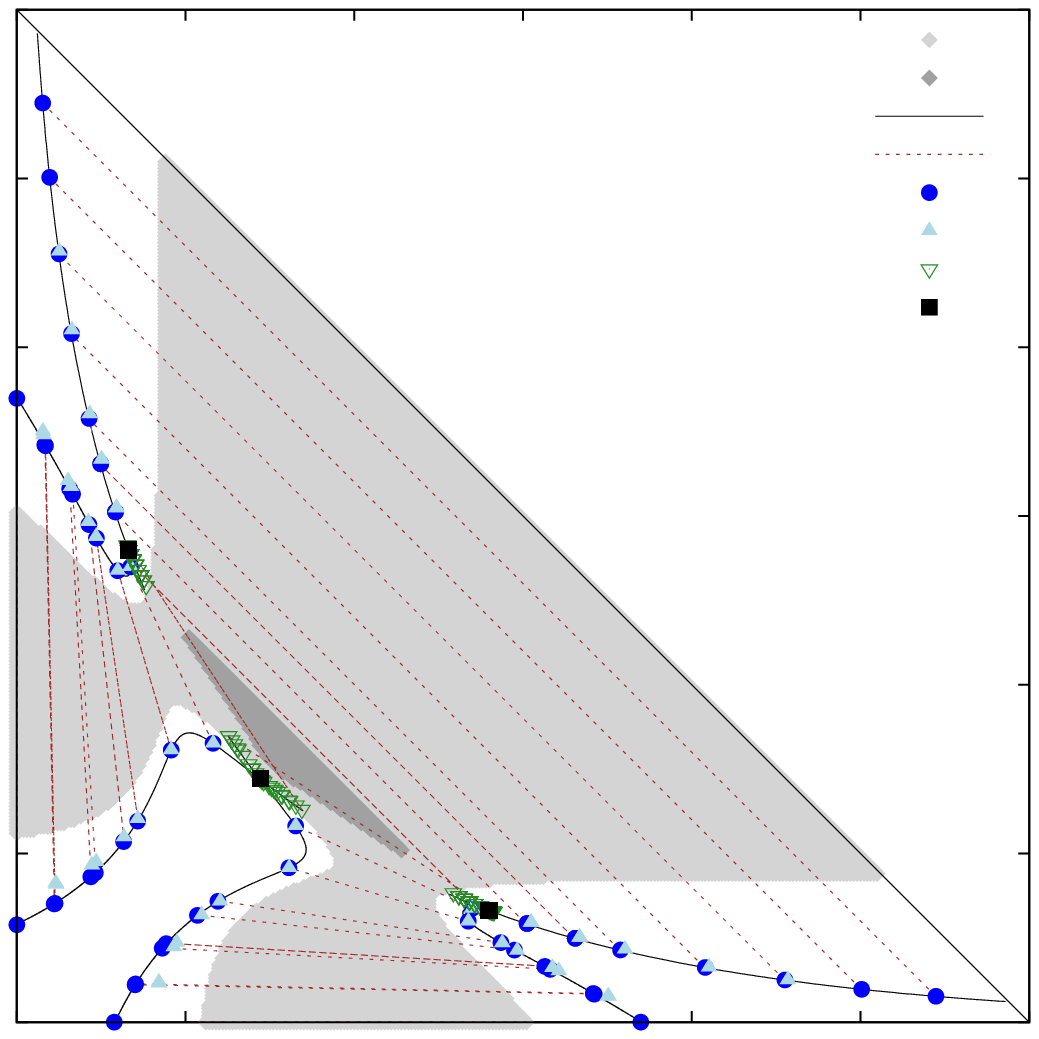}}
\label{graphBaselineFull}}
\hfill
\subfloat[][]{\resizebox{\columnwidth}{!}{\input{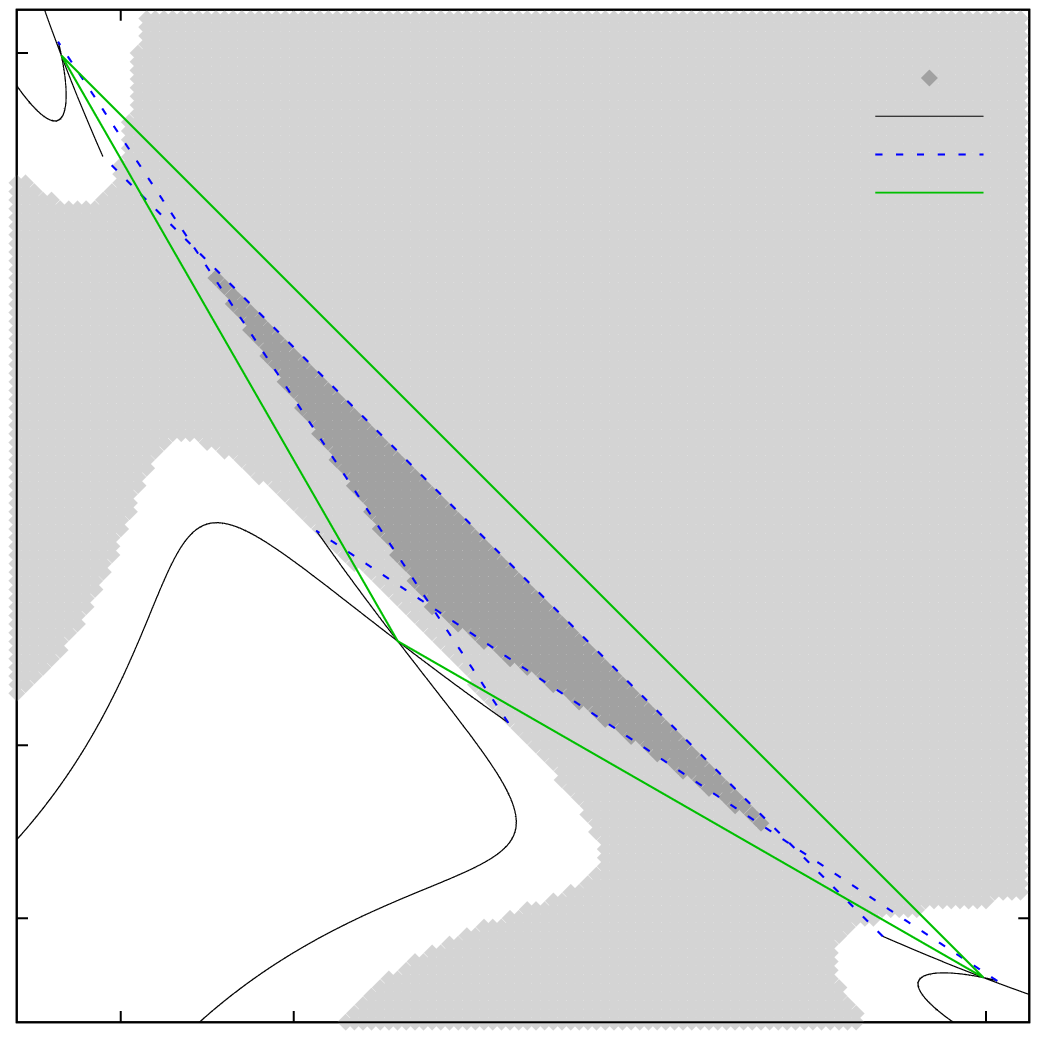}}
\label{graphBaseline3PhaseRegionZoom}}
\\
\subfloat[][]{\resizebox{\columnwidth}{!}{\input{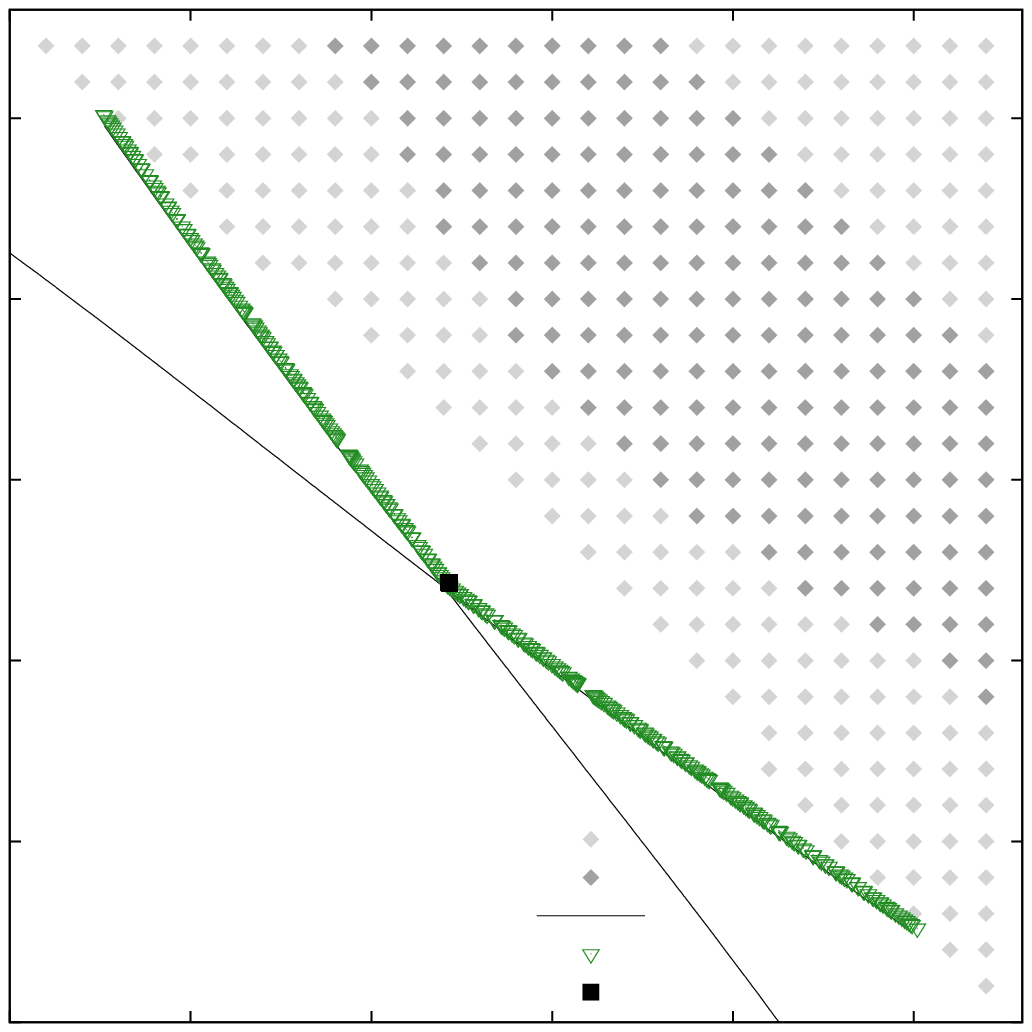}}
\label{graphBaseline3PhaseVaporZoom}}
\hfill
\subfloat[][]{\resizebox{\columnwidth}{!}{\input{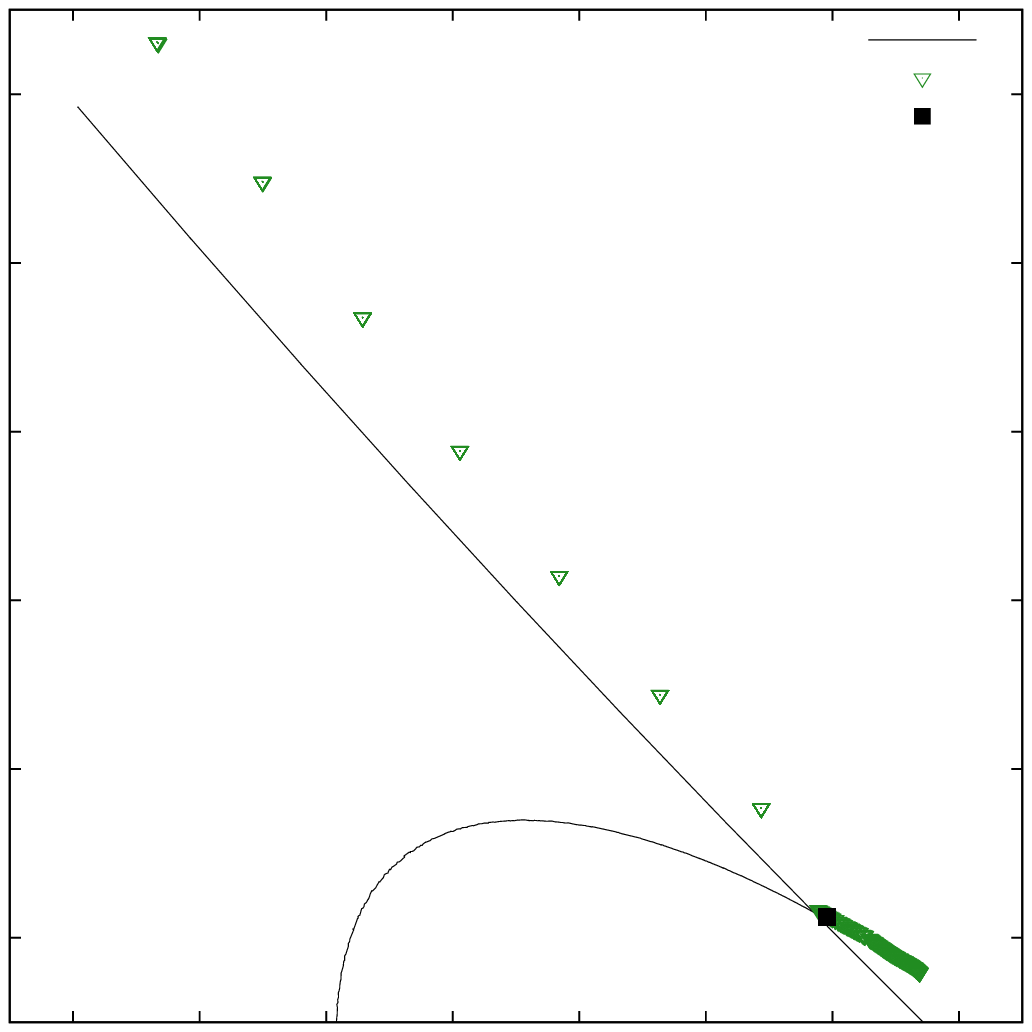}}
\label{graphBaseline3PhaseLiquidZoom}}
\caption{A baseline case of 3-phase behavior in a Type III-H phase diagram.  Part \protect\subref{graphBaselineFull} shows a phase diagram for a mixture identical to the  component in Figure \ref{graphTwoIdenticalComponents} except $\nu = 0.5$ $(\xi=0.0, \zeta=0.0, \Lambda=0.5)$.  We also include a comparison of the LB results using the two kinds of chemical potential gradient forcing.  Part \protect\subref{graphBaseline3PhaseRegionZoom} zooms in to show the three binodal intersections that define the full three-phase region and the connections between binodals that are used to inscribe the unconditionally unstable 3-phase region.  The minimization algorithm predicts 3-phase behavior for every point in this region, and the behavior was reflected in all LB simulations in this region, one example of this was shown in Figure \ref{graphBaselineNa008Nb008}.  In \protect\subref{graphBaseline3PhaseVaporZoom} we see the crossing binodals in the vapor region of the 3-phase behavior.  The LB simulations of metastable points follow the binodals well after the crossing point.  The LB simulations for three-phase points show a very small deviation from the point of intersection of the binodals, which is an error of $\sim 10^{-3}$.  Finally, part \protect\subref{graphBaseline3PhaseLiquidZoom} shows the crossing binodals in the A-rich liquid region of the 3-phase behavior; those in the B-rich liquid region are similar.  The LB simulations of metastable points follow the binodals well, but they show the same small deviation from the binodal intersection point as noted in \protect\subref{graphBaseline3PhaseVaporZoom}.}
\label{graphBaselineCase}
\end{figure*}

Proceeding from the neutral interaction case in Figure \ref{graphTwoIdenticalComponents}, we induced repulsive behavior between the two components by reducing the interaction parameter $\nu$.   Figure \ref{graphModerateRepulsive} shows this behavior with $\nu = 0.7$. The liquid-gas densities now depend on the concentration of the A and B components.  More mixed fluids show a gas density that is increased whereas the liquid density at coexistence is decreased. Also the composition of the liquid and gas are no longer equal. This has important consequences for the phase separation dynamics. Such a system will first phase-separate into liquid and gas phases of approximately equal composition in a process dominated by hydrodynamics, and then the domains will slowly exchange components through diffusion until the final equilibrium compositions are reached. At high densities a further miscibilty gap appears showing liquid-liquid phase separation where the two VDW fluids behave as a binary liquid.  The lattice Boltzmann simulations recover the predicted phase-behavior well.

For attractive inter-component interactions $\nu>1$ we find the opposite behavior and the difference between liquid and gas densities increases for mixed components. An example of this (for a slightly asymmetric mixture) is shown in Figure \ref{graphAttractiveAboveBothCriticalPoints}.

\begin{figure}
\centering
\resizebox{\columnwidth}{!}{\input{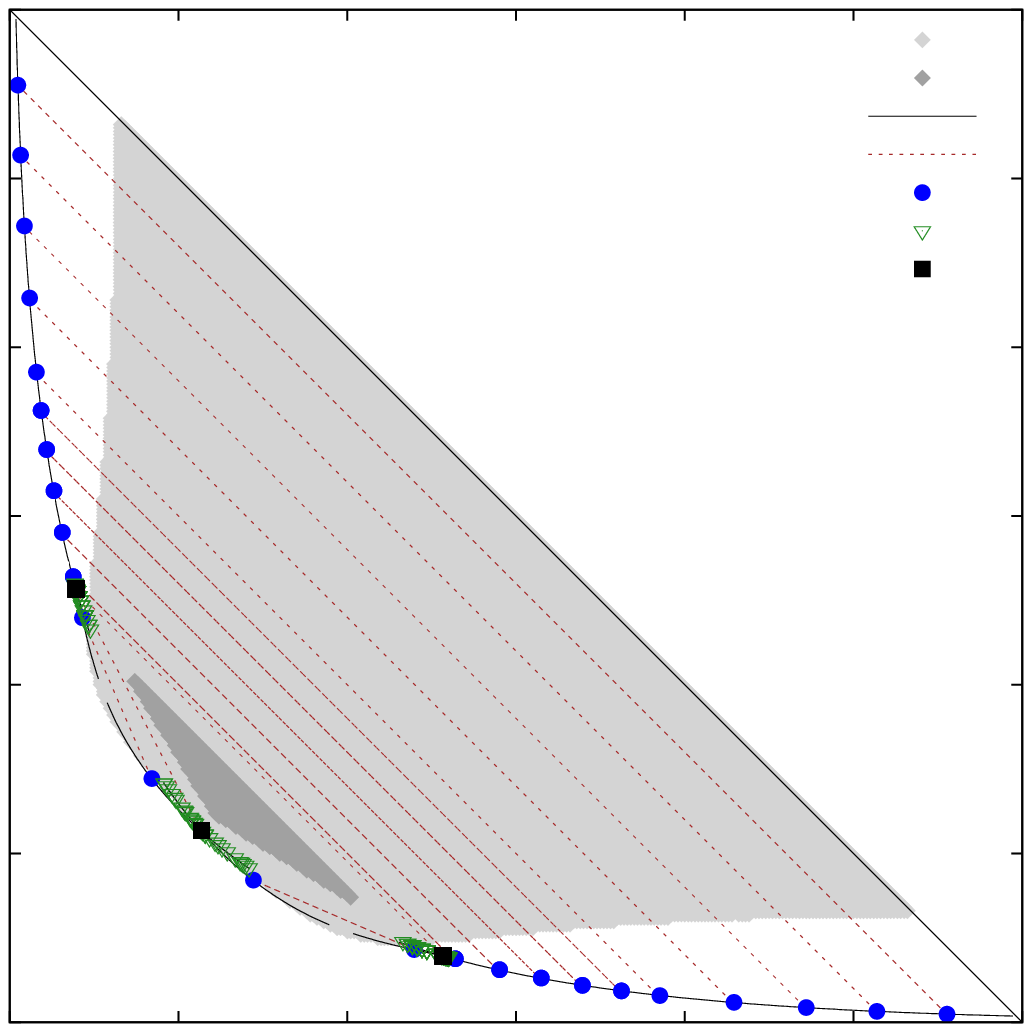}}
\caption{A Type III-H phase diagram showing a variation on Figure \ref{graphBaselineCase} where the lattice temperature is above the critical temperatures of each component, but the interaction parameter is repulsive enough to still elicit three-phase behavior.  It was generated using parameters of $\theta_{cr}^A = \theta_{cr}^B = 0.32$, $\rho_{cr}^B = 1.0$, and $\nu = 0.2$ $(\xi=0.0, \zeta=0.0, \Lambda=0.8)$.  The maximum density ratio is almost $105$.}
\label{graphRepulsiveAboveBothCriticalPoints}
\end{figure}

Gradually decreasing $\nu$ further to $\nu = 0.5$  leads to a pinching of the liquid-gas binodals and the creation of two critical points. At the same time the liquid-liquid phase separation region expands. Then the two separated liquid-gas binodal get close to the liquid-liquid binodal. Before they merge, however, the liquid-liquid critical point splits into two liquid-gas critical points and the binodals of the two new liquid-gas binodals and the liquid-liquid binodal meet at three-phase coexistence points, similar to the phase diagram of Figure \ref{graphRepulsiveAboveBothCriticalPoints}. The liquid-gas critical points then approach each other and merge.  In the creation of the three-phase region the three-phase coexistence points leads to binodals from this binary liquid region that intersect the liquid-gas coexistence curves as shown in Figure \ref{graphBaselineCase}. This complex transition as a function of $\nu$ is shown in movie 1 in the supplemental material.

The binodal lines that define the vapor densities for the two liquid-vapor regions intersect at the gas-phase of the three phase coexistence.  These three binodal intersections now define a new region in the phase diagram that exhibits either metastable 2-phase behavior extending the now meta-stable binodals or 3-phase behavior.  Initial density pairs that always minimize to three phases in this region are indicated by the dark grey regions of our phase diagrams. Intuitively it is clear that the meta-stable binodals have to end when one branch intersects with the spinodal region, because this branch now has to undergo a second round of phase separation leading to three-phase behavior.

The minimization algorithm predicts 3-phase behavior for every point in this region, and the behavior was reflected in all LB simulations in this region.  The LB simulations of metastable points follow the binodals well after the crossing point as shown in Figure \ref{graphBaselineCase} (c) and (d).  The LB simulations for three-phase points show only very small deviation from the point of intersection of the binodals.

We exhaustively tested every density pair in the full three-phase region with LB simulations, and as seen in Figure \ref{graphBaselineFull} the LB simulations recover both the metastable behavior and 3-phase behavior within the three-phase region quite well.  In particular, every simulation within the unconditionally unstable three-phase region (see Figure \ref{graphBaseline3PhaseRegionZoom}) exhibited 3-phase behavior.  Metastable points that were initialized with three phases held the 3-phase behavior as well as points in the unconditionally unstable three-phase region.  Metastable points that were initialized with two phases also held the 2-phase behavior very well and followed the theoretical binodals after the points of intersection.  However, we note that all 2-phase metastable LB simulations show small inaccuracies that we anticipate may be attributed to interfacial effects, but the analysis of which are outside the scope of this paper.

We used the baseline mixture in Figure \ref{graphBaselineCase} to perform a comparison between the two forcing methods based on chemical potential gradients.  This comparison was performed only in the two-phase regions of the phase diagram.  We found that the ``nid" chemical potential forcing method (Eq. \ref{forceMuId}) was greatly outperformed - both in terms of accuracy and stability - by the ``log" method of Eq. (\ref{forceMuLog}).  Given this, we based all subsequent simulations on only the method of Eq. (\ref{forceMuLog}).

Figure \ref{graphRepulsiveAboveBothCriticalPoints} is an example of a symmetric mixture where the lattice temperature is below the common critical temperature of the two components. For this mixture both components severely repel each other ($\gamma=0.2$).  The behavior shown in this particular phase diagram is striking: despite the fact that we are well above the critical temperature of either mixture this phase diagram shows two separate symmetric regions of liquid-gas phase separation as well as a three-phase coexistence. This is particularly unexpected since we previously observed in Figure \ref{graphModerateRepulsive} that liquid-gas phase separation was suppressed for $\gamma<1$. For this peculiar kind of liquid-gas phase separation the gas and liquid have substantially different compositions and the actual density of one of the components is larger in the gas than in the liquid. 
In terms of differential geometry Korteweg identified these additional liquid-gas regions with ``accessory plaits" to the free energy surface \cite{Sengers2002}. Encouragingly the lattice Boltzmann simulations are able to recover the predicted phase separation behavior well. Note that this behavior was also shown in movie 1 in the supplemental material.

\begin{figure}
\centering
\resizebox{\columnwidth}{!}{\input{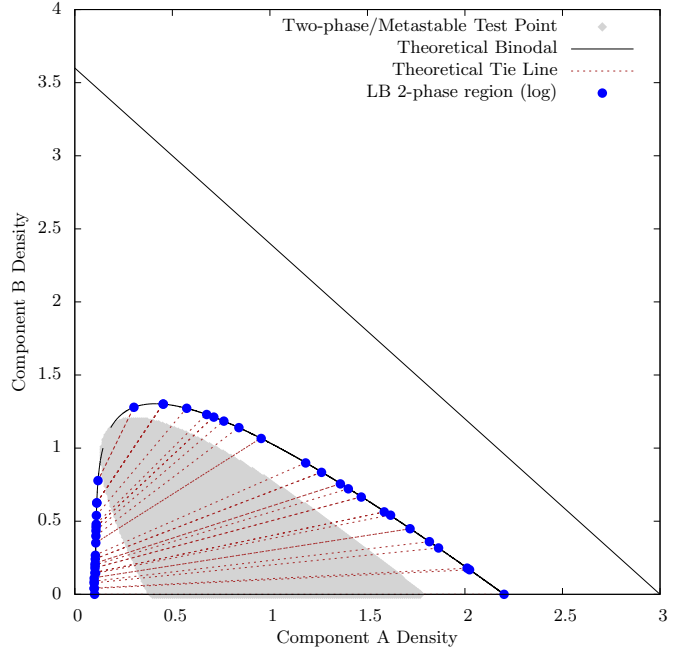}}
\caption{A Type II phase diagram showing a lattice temperature that is in between the critical temperatures of the two components ($\theta_{cr}^B < \theta < \theta_{cr}^A$) and asymmetric critical densities ($\rho_{cr}^A < \rho_{cr}^B$); the interaction parameter is neutral.  It was generated using parameters of $\theta_{cr}^A = 0.5$, $\theta_{cr}^B = 0.3$, $\rho_{cr}^B = 1.2$, and $\nu = 1.0$ $(\xi=0.090909, \zeta=0.162791, \Lambda=0.013339)$.  Six test points defaulted to values for $\kappa^{cc'}$ derived from single-component theory; the values were all between 3.6 and (slightly above) 3.7.}
\label{graphNeutralAsymmetric}
\end{figure}

\subsubsection{Asymmetric Components}
So far our analysis has focused on symmetric mixtures, corresponding to points in the global phase diagram on the $\zeta=0$ axis of Figure \ref{globalPhaseDiagram}. However, there is nothing in our lattice Boltmann method that requires this choice. In the following we show a few example of simulations for $\zeta\neq 0$.

Figure \ref{graphNeutralAsymmetric} shows the phase diagram for a mixture of two components with asymmetric critical traits.  The A- and B-components are asymmetrical in both their critical temperatures and critical densities, but they have a neutral interaction between them ($\nu=1$).  The overall appearance of this phase diagram is similar to that of a single-component VDW fluid with the B-density taking over the role of the temperature. The LB simulations recovered the theoretical expectations well overall; however, the accuracy of the vapor density results fell off by an order of magnitude ($\sim 10^{-3}$ error) at the higher density ratios. Such systems are of interest in that a change in composition can act in a similar way to a change in temperature to induce phase separation. This is the basic phenomenon in the formation of assymetric precipitation membranes\cite{akthakul2005lattice}.

\begin{figure}
\centering
\resizebox{\columnwidth}{!}{\input{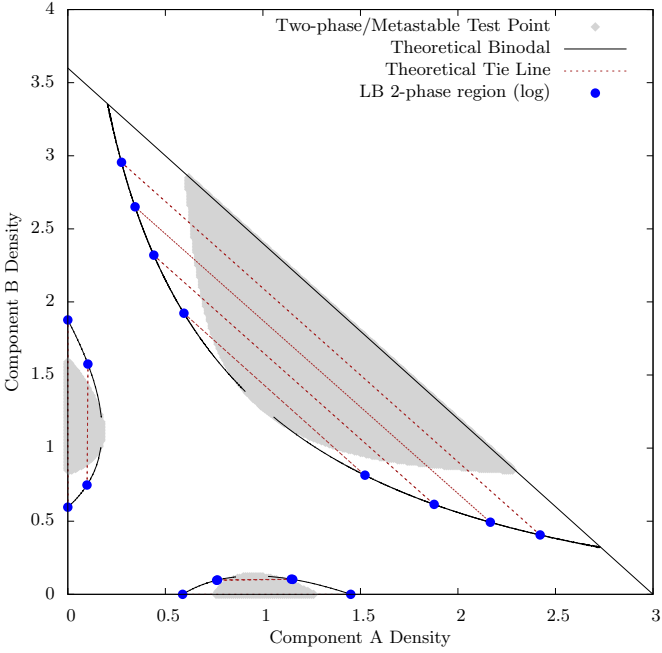}}
\caption{A Type III-H phase diagram showing a lattice temperature that is close enough to the critical temperatures of each component that there are three distinct regions of phase separation; the interaction parameter is repulsive.  It was generated using parameters of $\theta_{cr}^A = 0.35$, $\theta_{cr}^B = 0.36$, $\rho_{cr}^B = 1.2$, and $\nu = 0.6$ $(\xi=-0.090909, \zeta=0.104859, \Lambda=0.403308)$.}
\label{graphRepulsiveNearCriticalPoint}
\end{figure}

Figure \ref{graphRepulsiveNearCriticalPoint} is another mixture of components that are asymmetric in both critical temperatures and densities, ($\zeta\neq 0, \xi\neq 0$). In this case, the lattice temperature is slightly below the critical temperatures of each component, and the cross-component interaction is moderately repulsive.  The net result is three separate domains with large gaps separating them: a liquid-vapor region rich in A-component, a liquid-vapor region rich in B-component, and a binary liquid region.  This particular example is similar to what the mixture in Figure \ref{graphModerateRepulsive} would show at a higher temperature (except for the slight asymmetry). 

\begin{figure}
\begin{center}
\resizebox{\columnwidth}{!}{\input{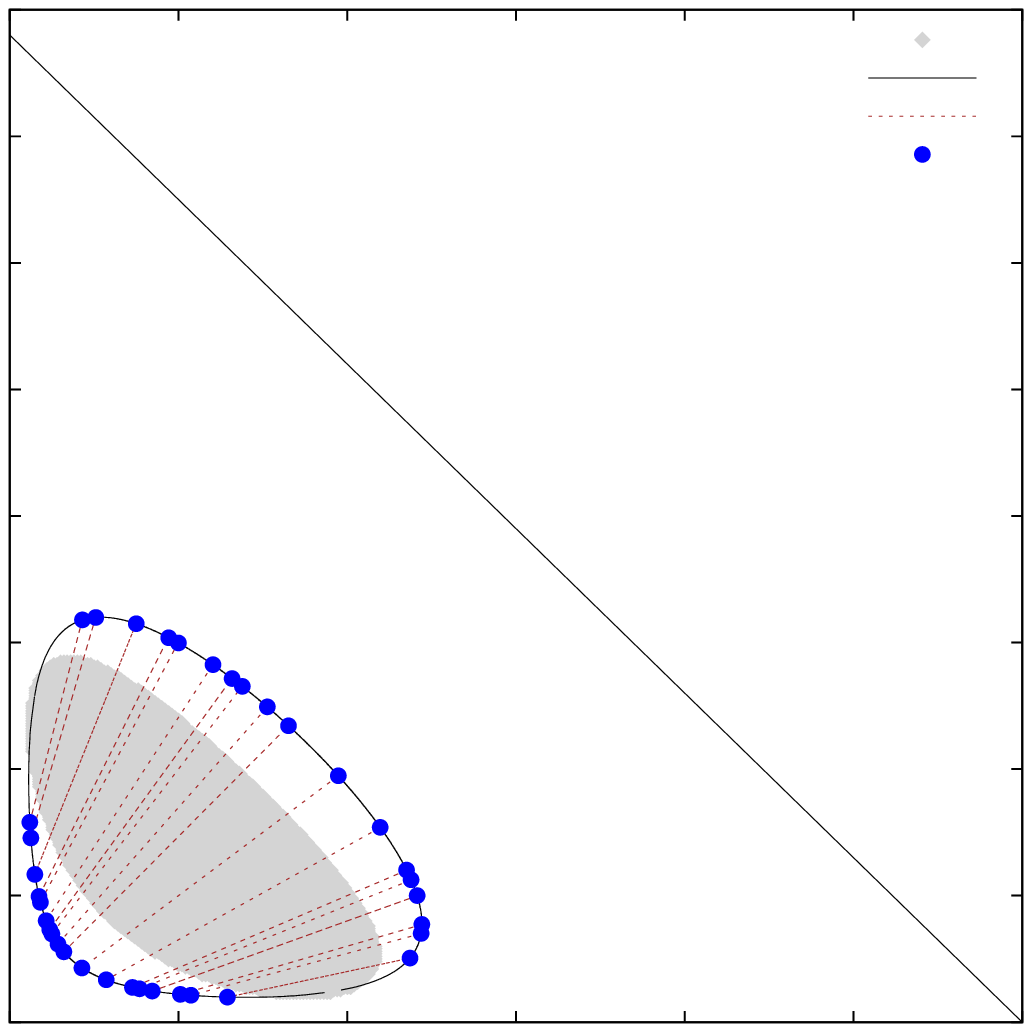}}
\end{center}
\caption{This phase diagram shows a mixture with a lattice temperature that is above the asymmetric critical temperatures of the individual components ($\theta_{cr}^A < \theta_{cr}^B < \theta$); however, the interaction parameter is attractive enough to still produce two-phase behavior.  It was generated using parameters of $\theta_{cr}^A = 0.3$, $\theta_{cr}^B = 0.31$, $\rho_{cr}^B = 1.3$, and $\nu = 1.5$ $(\xi=-0.130435, \zeta=0.146515, \Lambda=-0.483813)$.}
\label{graphAttractiveAboveBothCriticalPoints}
\end{figure}

Figure \ref{graphAttractiveAboveBothCriticalPoints} illustrates the behavior when asymmetric components that have a moderate affinity for each other are mixed.  The lattice temperature is again below the critical temperatures of both components, which has the effect of detaching the spinodal region from the pure component axes and creating a ``bubble" of 2-phase behavior bordered by two critical points. Such a system will have the unusual property that two pure gases above their critical temperatures will, when mixed, phase-separate into a liquid and a gas phase.

\begin{figure}
\centering
\resizebox{\columnwidth}{!}{\input{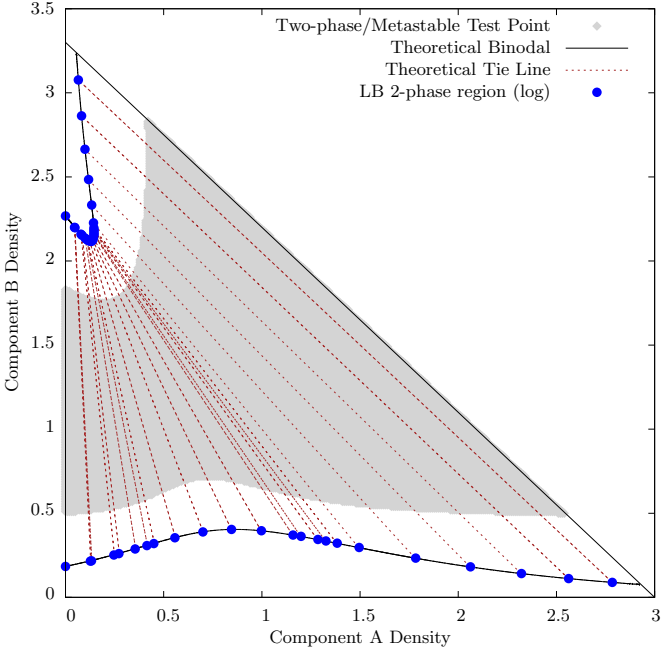}}
\caption{A mixture with a lattice temperature that is above the critical temperature of one component but below that of the other ($\theta_{cr}^A < \theta < \theta_{cr}^B  $) and with repulsive interactions between the species.  It was generated using parameters of $\theta_{cr}^A = 0.32$, $\theta_{cr}^B = 0.45$, $\rho_{cr}^B = 1.1$, and $\nu = 0.5$ $(\xi=-0.047619, \zeta=0.214724, \Lambda=-0.511663)$.}
\label{graphRepulsiveAboveOneCriticalPoint}
\end{figure}

If one of the mixtures is well above its critical point and the other one below it and there is a significant repulsive interaction we find a merged liquid-gas liquid-liquid binodal. Such a system is shown in Figure \ref{graphRepulsiveAboveOneCriticalPoint}. In this case we have only one binodal, and no critical point, similar to the neutral case shown in Figure \ref{graphTwoIdenticalComponents}. It is interesting to note that in Figure \ref{graphRepulsiveAboveOneCriticalPoint} there is a continuous transition between liquid-gas and liquid-liquid coexistence. Only the sharp turn in the binodal around $(\rho^A,\rho^B)\approx (0.2,2.2)$ gives a soft indication where the change in slope of the tie-lines changes from $-45^\circ$ indicating equal density for the two phases of a liquid-liquid coexistence to slope of $-90^\circ$, indicating a larger density difference indicative of liquid-gas coexistence. This shows that for binary mixtures there is no clear distinction between liquid-liquid and liquid-gas phase-separation.

\begin{figure}
\captionsetup[subfloat]{captionskip=10pt}
\centering
\subfloat[][]{\resizebox{\columnwidth}{!}{\input{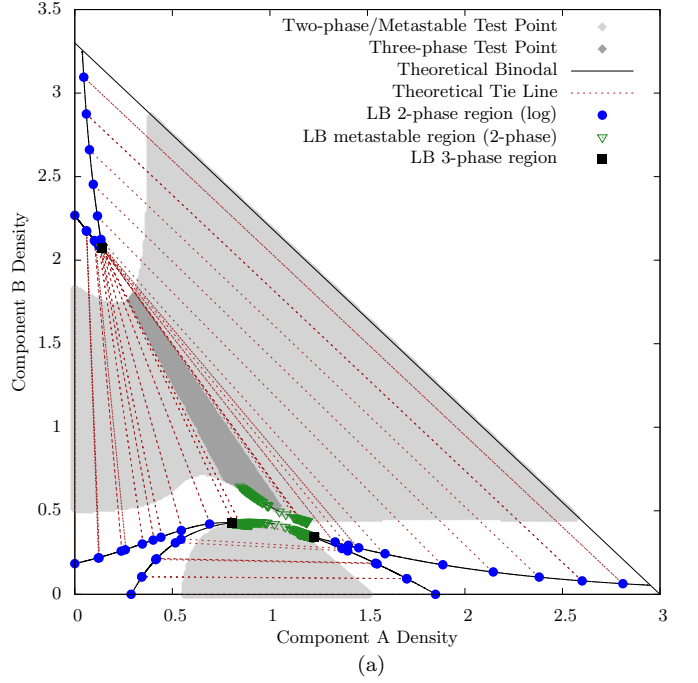}}
\label{graphRepulsiveAsymmetricFull}}\\
\subfloat[][]{\resizebox{\columnwidth}{!}{\input{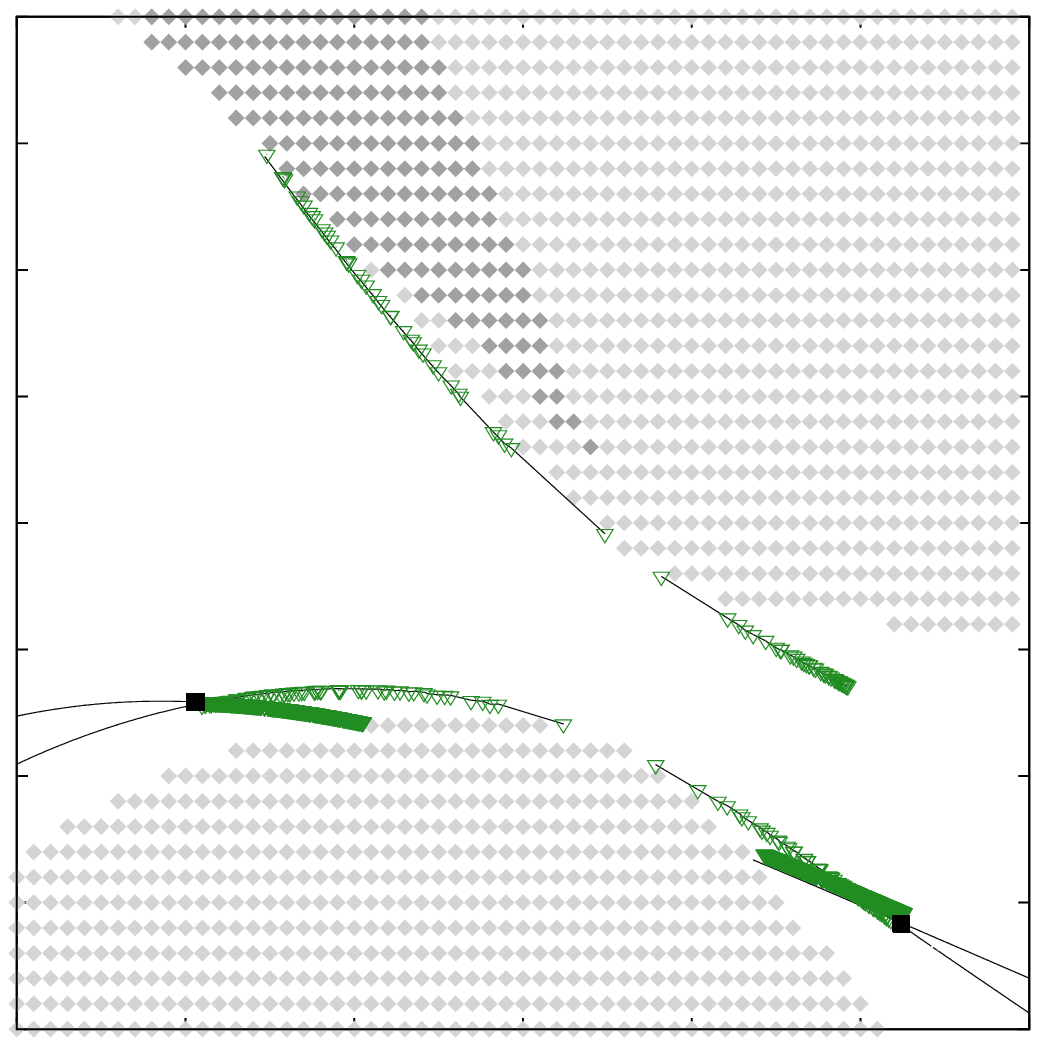}}
\label{graphRepulsiveAsymmetricZoom}}
\caption{Phase diagram with asymmetries that have produced an auxiliary binodal for the metastable region of the phase diagram.  A zoomed in view of the auxiliary binodals is shown in \protect\subref{graphRepulsiveAsymmetricZoom}.  This phase diagram was generated using parameters of $\theta_{cr}^A = 0.4$, $\theta_{cr}^B = 0.45$, $\rho_{cr}^B = 1.1$, and $\nu = 0.5$ ($\xi=-0.047619$, $\zeta=0.106145$, $\Lambda=0.502825$).  Unlike the rest of the phase diagrams shown, the metastable and 3-phase LB simulations were done with $\kappa^{cc'} = 0.2$.  One point (0.0,1.0) used the single-component value of $\kappa^{cc'} = 2$.}
\label{graphRepulsiveAsymmetric}
\end{figure}

In Figure \ref{graphRepulsiveAsymmetric}, we see perhaps our most interesting phase diagram with effects arising due to the asymmetry of the components.  Relative to our very symmetric, well-behaved basesline case in Figure \ref{graphBaselineCase}, only two parameter adjustments were made: the critical temperature of the B-component was raised to 0.45 (from 0.40) and the critical density of the B-component was raised to 1.1 (from 1.0).  All other parameters are unchanged. For symmetric mixtures the two liquid-gas critical points have to merge with the liquid-liquid binodal, and that can only happen if the liquid-liquid binodal first generates a three-phase region with two critical points through the process described in the discussion of Figure \ref{graphBaseline3PhaseLiquidZoom}. For very asymmetric systems this is not necessary. If the system is well above the critical point for one pure system the liquid-liquid critical point can merge with the other liquid-gas critical point without first forming a three-phase region (not shown). For only slightly asymmetric systems, however, the formation of a three-phase region inside one of the binodals is typical.

Now let us consider the specific example of Figure \ref{graphRepulsiveAsymmetric}. At first look, it appears that the binodal from the B-vapor simply rides up and over an independent A-liquid-vapor region to join the A-binary liquid binodal.  However, when zooming in to the peak of the A-liquid-vapor region, we see the situation is far more complicated.  The binodals cross again to define two points of the three-phase region. However in the the three-phase region a new, and to us completely unexpected, binodal for metastable two-phase behavior emerges.  This gap has the effect of dividing part of the metastable points into two new accessory plaits: one region at the apex of the A-liquid-vapor area and another at the bottom of the three phase region that is defined by two new, short binodal line segments. This binodal line is particularly unusual as it does not continuously connect to any of the three basic liquid-gas or liquid-liquid binodals.   The miscibility gap within the 3-phase region is certainly possible in the context of Korteweg's work (this phenomenon was demonstrated for a symmetrical case); however, the formation of accessory plaits that are encapsulated within the 3-phase region was entirely unexpected, and we were unable to find any references to this phenomenon in the literature \cite{Sengers2002, SengersLevelt2002}. At this point we have not been able to find another example of this specific behavior. 

\begin{figure}
\centering
\resizebox{\columnwidth}{!}{\input{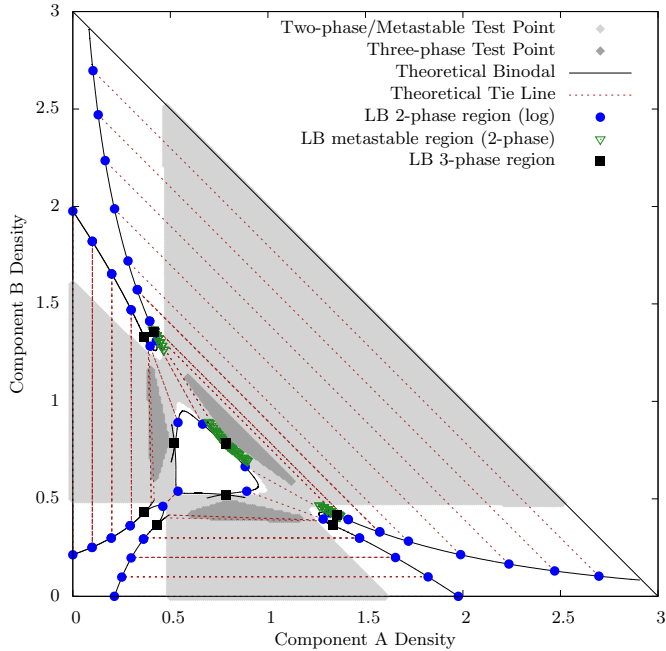}}
\caption{A phase diagram from the shield region showing three independent 3-phase regions shortly after the accessory plaits from each region connect.  The single-phase region in the middle is fully enclosed and contains binodal segments that are very nearly continuous.  The intermediary 2-phase regions contain points where metastable behavior would not be anticipated.  This phase diagram was generated using parameters of $\theta_{cr}^A = \theta_{cr}^B = 0.427$, $\rho_{cr}^B = 1.0$, and $\nu = 0.5636$ $(\xi=0.0, \zeta=0.0, \Lambda=0.4364)$.  The metastable and 3-phase LB simulations were done with $\kappa^{cc'} = 0.2$, and we only show the metastable results associated with one 3-phase region to more clearly depict the binodal segments elsewhere.}
\label{graphShield1}
\end{figure}

\subsubsection{Shield Region}
The shield region depicted in the upper, center of Figure \ref{globalPhaseDiagram} encloses a zone where 4-phase behavior between two VDW fluids is theoretically possible.  The transition through this region was first described by Korteweg using the tools of differential geometry, and his phase diagrams were replicated by computational means following the work of Scott and van Konynenburg \cite{Sengers2002}.  One of our goals was to replicate this process using lattice Boltzmann and to obtain a stable LB simulation of 2-component, 4-phase behavior.  The general strategy was to use the center of the shield region as identified by \cite{vanKonynenburg1968, vanKonynenburg1980} ($\zeta = 0,\Lambda = 0.4364$) and gradually increase the critical temperatures of the components (i.e. a deeper quench).  Although we fell short of observing 4-phase behavior, we found that the simple D1Q3 model was still able to replicate the transition through the shield region well.  Note that the description and phase diagrams in this subsection are all for symmetric components.

Our exploration of the shield region is shown in movie 2 of the supplemental material.  We started by creating a phase diagram with $\theta_c=0.4$ for both the A- and B-components, which yielded a phase diagram remarkably similar to the case shown in Figure \ref{graphRepulsiveNearCriticalPoint}.  We then increased the critical temperatures in small, uniform increments (i.e. equal changes to $\theta_c^A, \theta_c^B$) to quench the mixture temperature even farther below the critical temperatures.  As the quench became deeper, each region of the phase diagram developed independent 3-phase behavior with associated metastable accessory plaits.  The critical points of each accessory plait eventually coincide at $\theta_{cr}=0.427$, and the phase diagram regions merge, which isolates a ``bubble" of single-phase behavior in the middle of the phase diagram.  This is shown in Figure \ref{graphShield1}.


When initialized in near-equilibrium profiles in the three-phase regions, the LB simulations hold the predicted densities for all three regions well.  The three-phase regions are separated by regions of 2-phase behavior that form shortly after the critical point merger.  The majority of these 2-phase points are not bound by a three-phase region and are thus not expected to exhibit metastable behavior, and the LB simulations show that the predicted tie lines are recovered.

\begin{figure*}
\captionsetup[subfloat]{captionskip=10pt}
\centering
\subfloat[][]{\resizebox{0.49\textwidth}{!}{\input{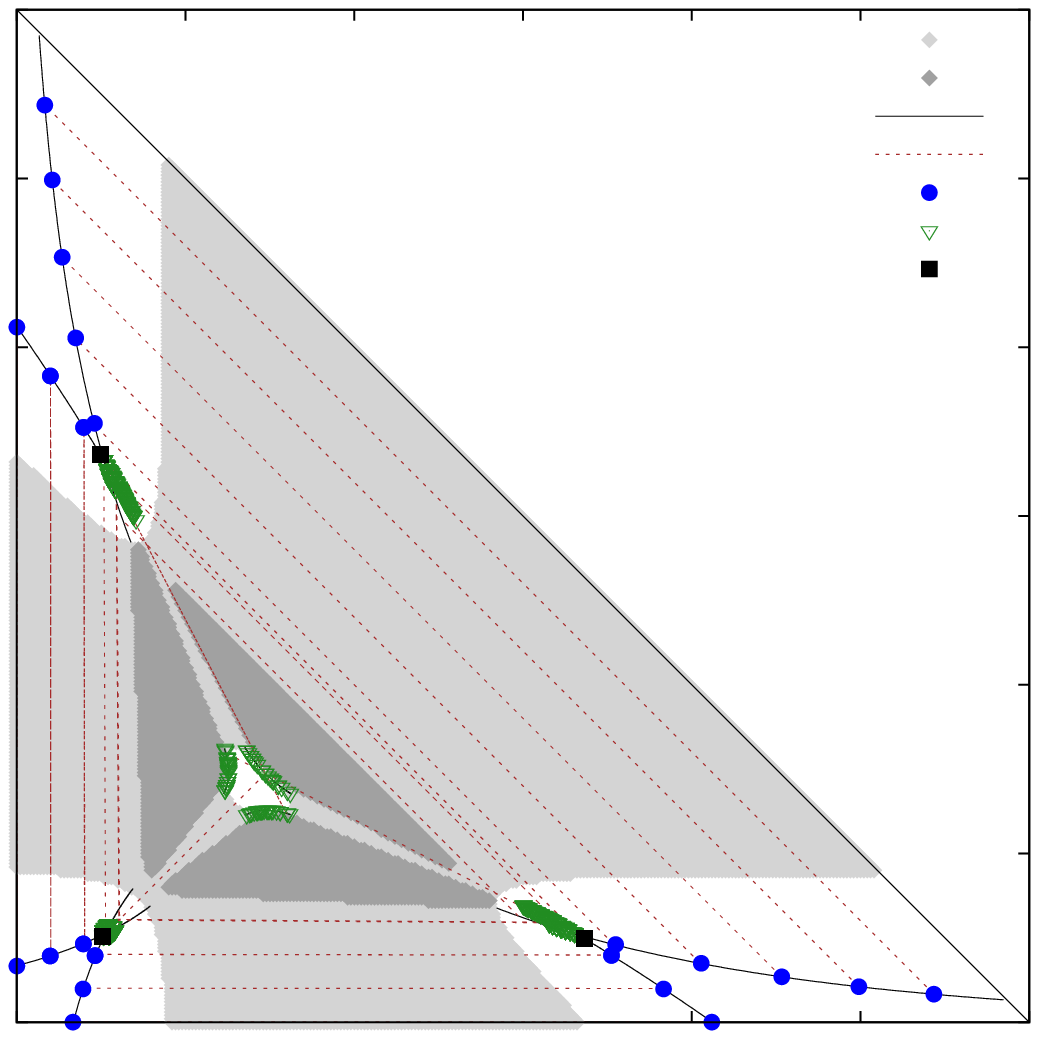}}
\label{graphShield2Full}}
\hfill
\subfloat[][]{\resizebox{0.49\textwidth}{!}{\input{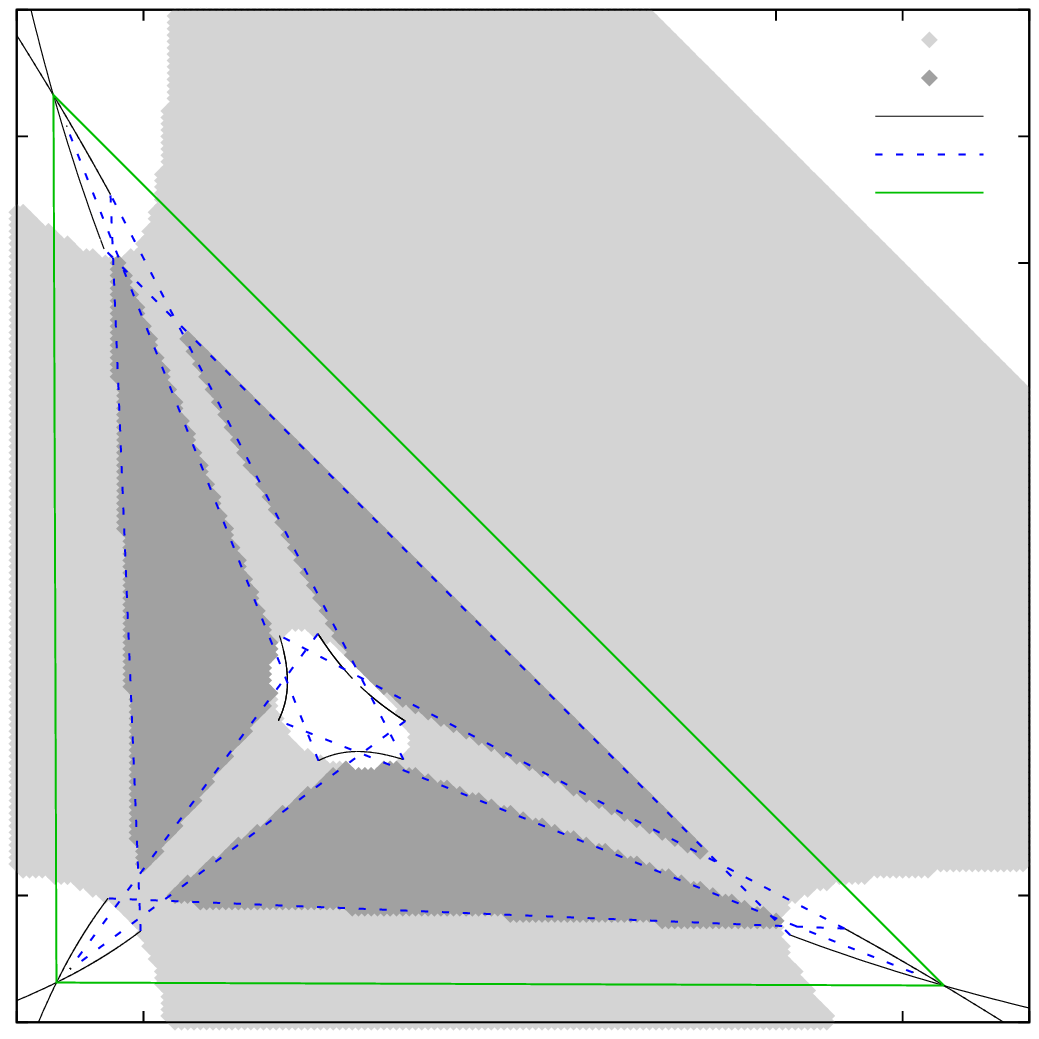}}
\label{graphShield2Zoom1}}
\\
\subfloat[][]{\resizebox{0.49\textwidth}{!}{\input{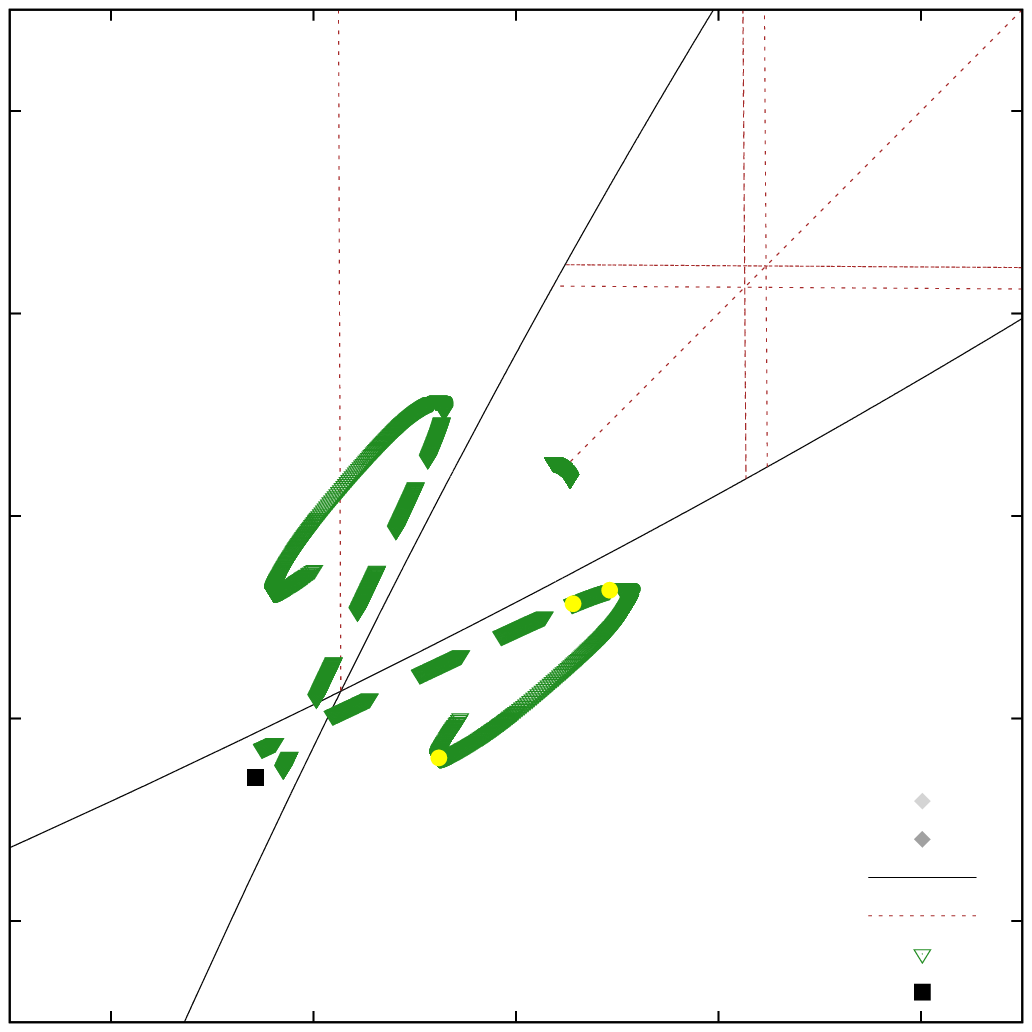}}
\label{graphShield2Zoom2}}
\hfill
\subfloat[][]{\resizebox{0.49\textwidth}{!}{\input{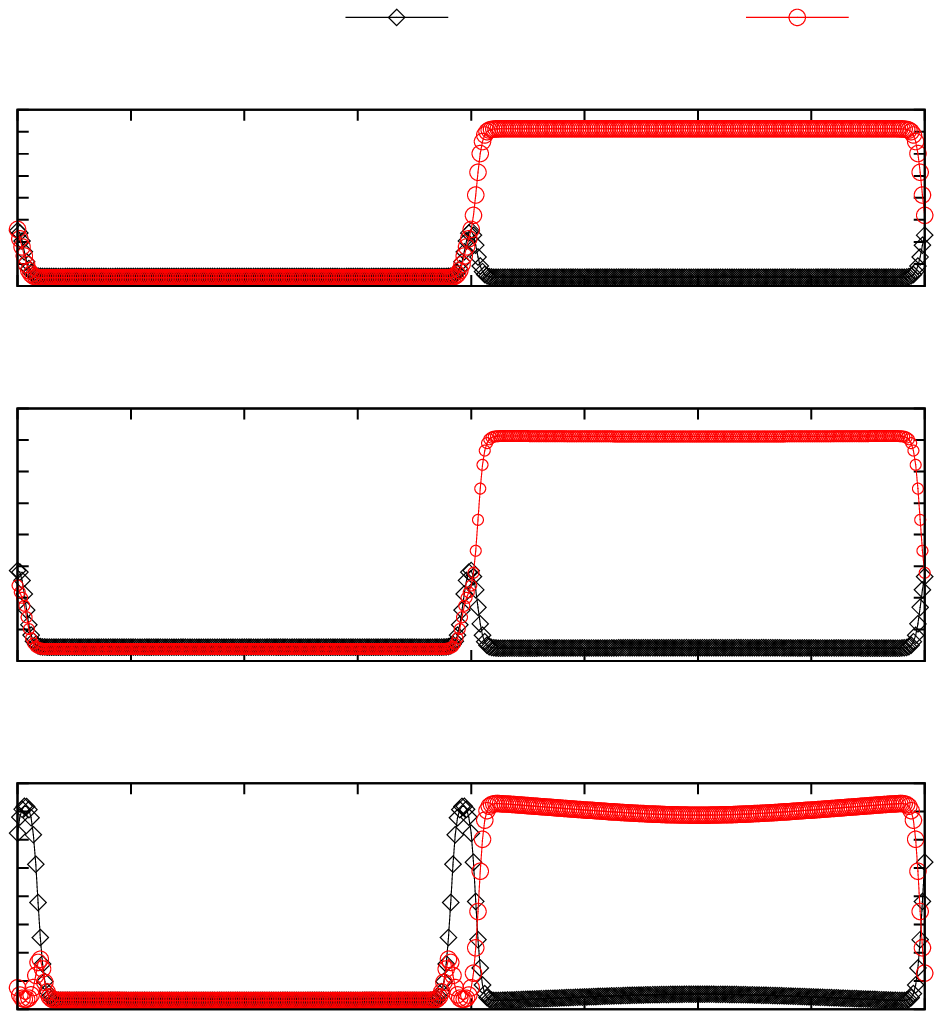}}
\label{graphShieldLB}}
\caption{A phase diagram from the shield region progressing towards 4-phase behavior.  Part \protect\subref{graphShield2Full} shows regions of unconditionally unstable 3-phase behavior, but they now share the same three equilibrium phases.  The intermediary 2-phase regions are now metastable.  Part \protect\subref{graphShield2Zoom1} shows binodals in the middle single-phase region separated into three segments.  Parameters are $\theta_{cr}^A = \theta_{cr}^B = 0.45$, $\rho_{cr}^B = 1.0$, and $\nu = 0.5636$ $(\xi=0.0, \zeta=0.0, \Lambda=0.4364)$.  Metastable and 3-phase LB simulations used $\kappa^{cc'} = 0.2$.  Part \protect\subref{graphShield2Zoom2} highlights the lower left of the 3-phase density of \protect\subref{graphShield2Full}. The metastable simulations lose track of their respective binodals.  Part \protect\subref{graphShieldLB} elucidates this behavior by showing a series of LB simulations that follow the curve of metastable results. Three yellow circle symbols in \protect\subref{graphShield2Zoom2} correspond to lattice site 100 shown in \protect\subref{graphShieldLB} (see text for details).}
\label{graphShield2}
\end{figure*}

Further increase of the critical temperatures brings us closer to the theoretical 4-phase behavior, and the first sign of this is when the densities associated with 3-phase behavior coalesce into a single set.  The independence of the separate three-phase regions is lost, yet they are still separated by ribbons of 2-phase metastable behavior.  The single-phase bubble shrinks in size, which slowly zeroes in on the expected density of the fourth phase.  Curiously, the tie lines that define the metastable 2-phase ribbons have endpoints that retreat closer to the 3-phase densities.  This is shown in better detail in Figure \ref{graphShield2Zoom2}.

In Figure \ref{graphShield2Zoom2}, we see that the metastable 2-phase tie line stops short of the 3-phase density (defined by the intersecting binodals) to define a new, tiny binodal segment.  The deviation of the LB 3-phase point is approximately $10^{-2}$, an order of magnitude larger than the rest of our examples.  This tiny binodal also has a very noticeable affect on the LB simulations of the metastable 2-phase points along the main binodal lines, which we outline in Figure \ref{graphShieldLB}.  Starting from the simulations near the binodal intersection, the LB simulations recover the expected metastable 2-phase behavior.  As the simulations reach abeam the tiny binodal segment we observe a deviation from the predicted binodal. The reason for this can be seen in Figure \ref{graphShieldLB}. Inside the interface of between the two metastable phases the third of the three 3-phase densities begins to emerge. Note that this is not a full nucleation event, but rather an augmentation of the interface.  Although the new domain doesn't fully form, it nonetheless has the effect of deflecting the LB results away from the binodal lines and back towards the binodal intersection.  This new effect was only seen in this simulation and prevented the LB simulations from recovering theoretical expectations for metastable behavior.

\begin{figure}
\centering
\resizebox{\columnwidth}{!}{\input{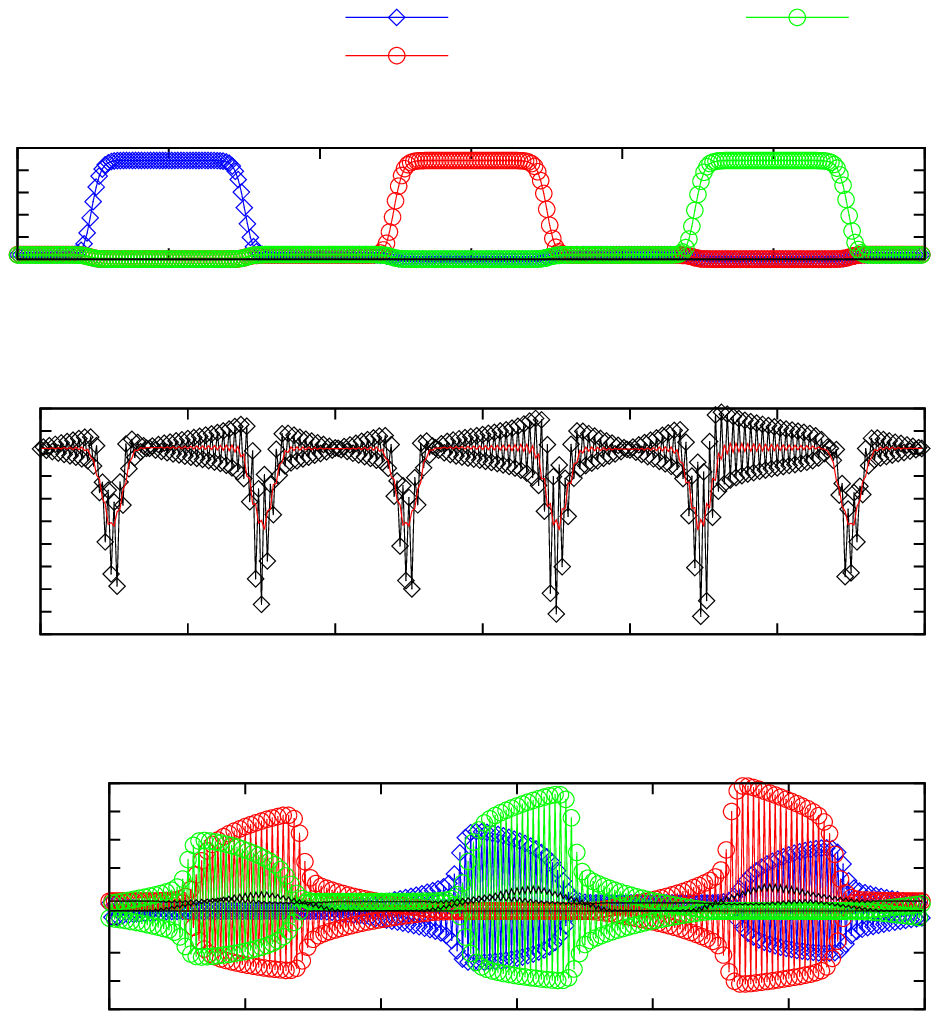}}
\caption{Simulation results for 3-components with the initial A,B,C components set (1.0, 1.0, 1.0), which exhibits thermodynamically consistent 4-phase equilibrium.  We use parameters $\theta^A_{cr} = \theta^B_{cr} = \theta^C_{cr} = 0.5$, $\rho_{cr}^B = \rho^C_{cr} = 1.0$, $\nu = 0.05$, and $\kappa^{cc'c''} = 2.0$.}
\label{graphLB3Components}
\end{figure}

\subsection{Three van der Waals Fluids}
To demonstrate the extensibility of our method, we implemented a LB simulation of a 3-component mixture of non-ideal fluids.  Since our LB implementation was designed to simulate each component in its own right rather than order parameter-style relationships, the extension was little more than a simple copy/paste operation in code.  Nothing else had to be derived for implementation, and we made zero changes to the corrections for thermodynamic consistency that we used in the two-component case.  

The LB simulation shown in Figure \ref{graphLB3Components} was initialized with a near-equilibrium density profile and parameters manually tuned to assure stability for at least 1,000,000 iterations.  The top view of the density profile shows a stable density ratio of $\sim 1700$.  The 2nd and 3rd panels are the pressure and chemical potentials, respectively.  Filtered values are plotted on top of the noisy raw profiles to show the bulk pressure is constant to $10^{-4}$ and constant chemical potentials to $10^{-5}$. This density ratio is at the limit of what can easily be achieved. We already see indications of instability for higher density ratios in the alternating oscillations in the chemical potential and pressure. Other simulations with lower equilibrium density ratios showed pressure and chemical potentials that were constant to $10^{-6}$ and did not show these oscillations.

\section{Outlook}
We have demonstrated that a LB method based on the minimization of a free energy function for a mixture of an arbitrary number of VDW fluids can recover the complex equilibrium behavior predicted for such a mixture.  The corrections to single-component simulations proposed earlier by Wagner \cite{Wagner2006} were applied to this method.  With these corrections applied, our method was shown to recover consistent and accurate thermodynamics across a wide range of symmetric and asymmetric two-component fluid mixtures.  We also demonstrated that it is very easily extended to simulate mixtures of three or more non-ideal fluid mixtures with equally consistent and accurate thermodynamic consistency.

Our discrete free energy was formulated in a manner reminiscent of the pseudopotential methods of Shan and Doolen \cite{Shan1995}.  This allowed us to identify the interaction strength $\psi$ in terms of other commonly used non-ideal interaction variables.  We also showed that this formulation can help reveal an appropriate choice of gradient stencil.  

The numerical stability of our simulations was greatly improved with only a basic application of the findings of Pooley and Wagner \cite{WagnerPooley2007}.  Using the common methods of implementing numerical parameters to tune interface and forcing strength led to phase separation with standard density ratios on the order of $\sim 20$.  But by ensuring the widths of phase interfaces in our initialized density profiles were at or above a minimum threshold, we were able to easily obtain density ratios over $150$.  Further manual optimization of our parameters combined with ensuring a minimum initial interface with resulted in a density ratio of over $1700$.  

In all cases, our LB simulation results recover all features of our phase diagrams very well.  Since our free energy minimization doesn't account for interface effects, the majority of our LB simulations do not lie exactly at the ends of the theoretical tie lines.  But most simulations show a $10^{-4}$ or less deviation from a binodal line after only 50,000 iterations.  Occasionally the error increases to $10^{-3}$, but allowing simulations to run past our iteration cap to reach full equilibrium shows that the error gradually shrinks as material diffuses among phases.

We were extremely pleased to learn that for such a simple model that included only three discrete lattice velocities and considered only bulk equilibrium properties, the LB simulations were able to replicate such a rich set of phase diagram features with outstanding accuracy.  Future extensions of this LB model will expand the method to higher dimensions, will examine the ability to recover a range of interfacial properties and, most importantly, the dynamics. We are particularly interested in extending this to evaporation phenomena, treated more phenomenologically in \cite{miller2014phase}.

\appendix

\section{Algorithm for generating phase diagrams \label{App1}} 

The theoretical phase diagrams by which we judged the performance of the LB method were created by numerically minimizing the underlying free energy.  A design decision was made to design a quasi-brute force minimization algorithm to accentuate the underlying physics of the mixture.  As this was a key component to the research, we provide a high-level description of the algorithm here; the C-code is open source and provided online \cite{ResearchCode2018}.

\begin{enumerate}
\item Loop over all ($A,B$) particle pairs below the line connecting van der Waals discontinuities for each component. $A$ is the number of particles of component A, and $B$ is the number of particles of component B.  \label{minimizationStep1}
\item Perform a stability analysis of the free energy at the point ($A,B$) via second derivatives with respect to component densities.
	\begin{enumerate}
	\item If the point is stable, phase separation is not expected.  Continue to the next ($A,B$) test point in Step \ref{minimizationStep1}.
	\item If the point is not stable, proceed with an attempt to divide the ($A,B$) particles among phases to minimize the free energy.  \label{minimizationEigenvector}
	\end{enumerate}
\item Initialize the free energy of the mixture, and choose an initial step size by which to vary the particle counts and volumina for each phase.
	\begin{enumerate}
	\item Assume equal volumina for the 3 allowed phases ($V1, V2, V3$).  For simplicity, we constrain the total volume of the system to equal 1, so each phase is initially allocated $1/3$.
	\item Use the eigenvector associated with the negative eigenvalue to divide the ($A,B$) particles between phases 1 ($A1, B1$) and 2 ($A2, B2$).  Phase 3 is initially empty. 
	\end{enumerate}	 
\item Create a 6x3 array of free energy trial values. \label{minimizationTrialArray}
	\begin{enumerate}
	\item For the 6 physical degrees of freedom (phase 1: $A1, B1, V1$; phase 2: $A2, B2, V2$), vary each independently by a positive, negative, and neutral step.
	\item Determine phase 3 ($A3, B3, V3$) by applying the conservation statements $N_A=A1+A2+A3$, $N_B=B1+B2+B3$, $V=V1+V2+V3$.	
	\end{enumerate}
\item Evaluate the free energy trial array to see if the minimum free energy in the array is less than that of the current particle/volume phase combinations.
	\begin{enumerate}
	\item If the the array has a new minimum free energy, declare that a phase change has occurred and save the associated particle/volume combination.  Keep the current step size and return to Step \ref{minimizationTrialArray} for the next iteration.
	\item If the minimum free energy is unchanged, halve the step size used to populate the free energy trial array and return to Step \ref{minimizationTrialArray} for the next iteration.
	\item Declare the free energy has been sufficiently minimized when the change in free energy is less than a chosen threshold (we use a threshold of $10^{-12}$).  Continue to Step \ref{minimizationPhaseDetector}.
	\end{enumerate}
\item Divide the particle counts for each component by the volumina of each phase to create the resulting densities of each phase.  Examine the densities that correspond to the minimum free energy to classify the resulting phase behavior. \label{minimizationPhaseDetector}
    \begin{enumerate}
    \item If there was no phase change, return to Step \ref{minimizationStep1} to evaluate the next (A,B) pair.
    \item If a stability analysis of the resulting densities shows a phase is still unstable, adjust the particles allocated to each phase to attempt another minimization.  We make this adjustment by packing the two stable phases together - which in this implementation have densities equal to $10^{-4}$ - into phase 1 and go to Step \ref{minimizationEigenvector} to split the unstable phase according to its unstable eigenvector for additional minimization iterations.
    \item If the phase change resulted in 2 or 3 stable phases, log the associated particle/volume data for use in creating the phase diagram for the mixture.  Go to Step \ref{minimizationStep1} to evaluate the next ($A,B$) particle pair.
    \end{enumerate}
\end{enumerate}

\bibliography{paper-bib,AW}

\end{document}